\documentclass{elsarticle}

\usepackage[rgb]{xcolor}

\usepackage[]{subfig}
\DeclareSubrefFormat{myparens}{#1(#2)}
\usepackage{caption}
\usepackage{moreverb, amsmath, amsfonts,amssymb,amsthm}
\usepackage{graphicx}
\usepackage{placeins}
\usepackage[ruled]{algorithm2e}
\usepackage{algorithmic}
\usepackage{booktabs}
\usepackage{multirow}
\usepackage{xspace}
\usepackage{hyperref}
\usepackage{todonotes}
\usepackage{geometry}

\usepackage{hyperref}
\journal{Journal of Parallel Computing}

\usepackage{adjustbox}
\usepackage{scalefnt}
\usepackage{tikz}
\usetikzlibrary{patterns,calc}
\usepackage{pgf}

\usetikzlibrary{positioning}
\usetikzlibrary{shapes.misc}

\makeatletter
\newenvironment{topbox}[2]{%
    \pgfmathsetlength\@tempdima{#2}%
    \pgfmathsetlength\pgf@yc{\pgfkeysvalueof{/pgf/inner ysep}}%
    \advance\@tempdima by -2\pgf@yc
    \begin{lrbox}{\@tempboxa}%
        \begin{minipage}[t]{#1}%
           \vspace{0pt}%
}{%
        \end{minipage}%
    \end{lrbox}%
    \ifdim\@tempdima>\dp\@tempboxa
        \dp\@tempboxa=\@tempdima
    \fi
    \box\@tempboxa
}
\makeatother

\tikzstyle{block} = [rectangle,fill=none,solid, draw=black, text centered,minimum width=2em, minimum height=2em,inner sep=0pt]
\DeclareRobustCommand\circled[1]{\tikz[baseline=(char.base)]{ \node[shape=circle, fill=white,draw,inner sep=0pt,text width= 1 em ,minimum size = 1 em, text centered] (char) {#1};}}
\tikzstyle{treenode} = [rectangle,fill=white,solid, draw=black, text centered,minimum size=1.5em, text width=1.5em,inner sep=0pt]
\DeclareRobustCommand\supernode[1]{\tikz[baseline=(char.base)]{ \node[shape=rectangle, fill=white,draw,inner sep=0pt,text width= 1 em ,minimum size = 1 em, text centered] (char) {#1};}}

\newcommand{\mapblock}[4]{
    \pgfmathtruncatemacro{\curproc}{ int(mod(#2-1,#3))*#4 + int(mod(#1-1,#4)) +1 };
}

\newcommand{\colorproc}[3]{
  \pgfmathsetmacro{\h}{#1 /(#2*#3)}
  \definecolor{proccolor}{hsb}{\h, 0.5, 0.8}
}

\newcommand{\blk}[6]{
\node[block,minimum width=#2em, minimum height=#3em #6] at (#4em,-#5em) {#1};
}

\newcommand{\dblknew}[4]{
\blk{#1}{#2}{#2}{#3}{#3}{#4}
}

\tikzset{cross/.style={cross out, draw=black, minimum size=1em, inner sep=0pt, outer sep=0pt,thick},
cross/.default={1em}}

\DeclareRobustCommand\lucross[0]{\tikz[baseline=(nbaseline.base)]{ \node[cross,thick] (char) {};\node (nbaseline) at ($(char) + (0em,-0.25em)$) {};}\xspace}
\DeclareRobustCommand\linvsquare[0]{\tikz[baseline=(nbaseline.base)]{ \node[rectangle,minimum size = 1em,draw=blue, thick] (char) {};\node (nbaseline) at ($(char) + (0em,-0.25em)$) {};}\xspace}

\DeclareRobustCommand\ainvfill[0]{\tikz[baseline= (nbaseline.base)]{ \node[circle,minimum size=.4em,fill=black] (char) {};\node (nbaseline) at ($(char) + (0em,-0.25em)$) {};}\xspace}

\newcommand{\mdblk}[6]{
    \pgfmathtruncatemacro{\x}{ int(mod(#2-1,#5))*#6 + int(mod(#2-1,#6)) +1 };
 \pgfmathsetmacro{\h}{\x /(#5*#6)}
\definecolor{currentcolor}{hsb}{\h, 0.5, 0.8}
\node[block,minimum width=#3 em, minimum height=#3 em,fill=currentcolor] at (#4 em,-#4 em) {\Proc{\x}};
}

\newcommand{\elblkFill}[5]{
\node[block,draw=none,minimum width=#2em, minimum height=#3em,fill=white,opacity=1] at (#4em,-#5em) {#1};
\draw[draw=black] ($(#4em,-#5em)- 0.5*(#2em,-#3em)$) -- ($(#4em,-#5em)- 0.5*(#2em,#3em)$);
\draw[draw=black] ($(#4em,-#5em) + 0.5*(#2em,#3em)$) -- ($(#4em,-#5em) + 0.5*(#2em,-#3em)$);
}
\newcommand{\eublkFill}[5]{
\node[block,draw=none,minimum width=#3em, minimum height=#2em,fill=white,opacity=1] at (#5em,-#4em) {#1};
\draw[draw=black] ($(#5em,-#4em)- 0.5*(#3em,-#2em)$) -- ($(#5em,-#4em)- 0.5*(-#3em,-#2em)$);
\draw[draw=black] ($(#5em,-#4em)- 0.5*(#3em,#2em)$) -- ($(#5em,-#4em)- 0.5*(-#3em,#2em)$);
}

\newcommand{\elublkFill}[5]{
	\elblkFill{#1}{#2}{#3}{#4}{#5}
	\eublkFill{#1}{#2}{#3}{#4}{#5}
}

\newcommand{\mlublk}[9]{
    \pgfmathtruncatemacro{\x}{ int(mod(#3-1,#8))*#9 + int(mod(#2-1,#9)) +1 };
 \pgfmathsetmacro{\h}{\x /(#8*#9)}
\definecolor{currentcolor}{hsb}{\h, 0.5, 0.8}
 \node[block,minimum width=#4em, minimum height=#5em,fill=currentcolor] at (#6em,-#7em) {\Proc{\x}};

    \pgfmathtruncatemacro{\x}{ int(mod(#2-1,#8))*#9 + int(mod(#3-1,#9)) +1 };
 \pgfmathsetmacro{\h}{\x /(#8*#9)}
\definecolor{currentcolor}{hsb}{\h, 0.5, 0.8}
 \node[block,minimum width=#5em, minimum height=#4em,fill=currentcolor] at (#7em,-#6em) {\Proc{\x}};
}
\newcommand{\mlblk}[9]{
    \pgfmathtruncatemacro{\x}{ int(mod(#3-1,#8))*#9 + int(mod(#2-1,#9)) +1 };
 \pgfmathsetmacro{\h}{\x /(#8*#9)}
\definecolor{currentcolor}{hsb}{\h, 0.5, 0.8}
 \node[block,minimum width=#4em, minimum height=#5em,fill=currentcolor] at (#6em,-#7em) {\Proc{\x}};
}
\newcommand{\mublk}[9]{
    \pgfmathtruncatemacro{\x}{ int(mod(#2-1,#8))*#9 + int(mod(#3-1,#9)) +1 };
 \pgfmathsetmacro{\h}{\x /(#8*#9)}
\definecolor{currentcolor}{hsb}{\h, 0.5, 0.8}
 \node[block,minimum width=#5em, minimum height=#4em,fill=currentcolor] at (#7em,-#6em) {\Proc{\x}};
}

\newcommand{\ignore}[1]{}

\newcommand{\Tr}{\ensuremath{\mathrm{Tr}}\xspace}

\newcommand{\CS}{\ensuremath{\mathcal C}\xspace}
\newcommand{\JS}{\ensuremath{\mathcal J}\xspace}
\newcommand{\IS}{\ensuremath{\mathcal I}\xspace}
\newcommand{\KS}{\ensuremath{\mathcal K}\xspace}

\newcommand{\CL}{\ensuremath{\CS_{L}}\xspace}
\newcommand{\CU}{\ensuremath{\CS_{U}}\xspace}

\newcommand{\pmat}{{PMatrix}\xspace}
\newcommand{\wt}[1]{\widetilde{#1}}
\newcommand{\Amat}{\ensuremath{A}\xspace}
\newcommand{\Lmat}{\ensuremath{L}\xspace}

\newcommand{\Proc}[1]{\ensuremath{P_{#1}}\xspace}

\newcommand{\etc}{\textit{etc}.\xspace}

\newcommand{\pexsi}{\texttt{PEXSI}\xspace}
\newcommand{\pselinv}{\texttt{PSelInv}\xspace}

\newcommand{\pardiso}{\texttt{PARDISO}\xspace}
\newcommand{\mumps}{\texttt{MUMPS}\xspace}
\newcommand{\superlu}{\texttt{SuperLU\_DIST}\xspace}

\newtheorem{theorem}{Theorem}

\begin{document}

%\category{G.4}{Mathematical Software}{}[Algorithm design and analysis]
%
%\category{G.4}{Mathematical Software}{}[Parallel and vector implementations]
%
%\category{I.1.2}{Symbolic and Algebraic Manipulation}{Algorithms}[Algebraic algorithms]
%
%\terms{Design, Performance}
%
%\keywords{selected inversion, sparse direct method, distributed memory
%parallel algorithm, high performance computation, electronic structure
%theory}

\begin{frontmatter}

\title{\texttt{PSelInv} -- A Distributed Memory Parallel Algorithm for
Selected Inversion : the non-symmetric Case} 


\author[lbl]{Mathias Jacquelin}
\ead{mjacquelin@lbl.gov}
\author[ucb,lbl]{Lin Lin}
\ead{linlin@math.berkeley.edu}
\author[lbl]{Chao Yang}
\ead{cyang@lbl.gov}

\address[lbl]{Computational Research Division, Lawrence Berkeley National Laboratory, Berkeley CA 94720 USA}
\address[ucb]{Department of Mathematics, University of California, Berkeley,  Berkeley CA 94720 USA}

\markboth{M. Jacquelin, L. Lin and C. Yang}{{\bf PSelInv} -- Parallel
Selected Inversion of a Sparse Non-Symmetric Matrix}

\begin{abstract}
This paper generalizes the parallel selected inversion algorithm called
\texttt{PSelInv} to sparse non-symmetric
matrices.  We assume a general sparse matrix $A$ has been decomposed as
$PAQ = LU$ on a distributed memory parallel machine, where $L,U$ are
lower and upper triangular matrices, and $P,Q$ are permutation matrices,
respectively.  The \texttt{PSelInv}
method computes selected elements of $A^{-1}$. The selection is confined by the sparsity pattern of the matrix
$A^{T}$. Our algorithm does not assume any symmetry properties of
$A$, and our parallel implementation is memory efficient, in the
sense that the computed elements of $A^{-T}$ overwrites the sparse
matrix $L+U$ \textit{in situ}.  \texttt{PSelInv} involves a large
number of collective data communication activities within different processor groups of
various sizes.  In order to minimize idle time and improve load
balancing, tree-based asynchronous communication is used to coordinate
all such collective communication.  Numerical results
demonstrate that \texttt{PSelInv} can scale efficiently to $6,400$
cores for a variety of matrices.
%We demonstrate the efficiency and accuracy of \texttt{PSelInv} by
%presenting a number of numerical experiments.  In particular, we show
%that \texttt{PSelInv} can run efficient on more than $XXX$ processors
%for modestly sized matrix arising from large-scale electronic
%structure calculations.
\end{abstract}


\end{frontmatter}

\section{Introduction}\label{sec:intro}

%\LL{asymmetric or non-symmetric? Need to unify}
%\MJ{I think non-symmetric is more commonly used so I have changed everything}

Let $A\in \mathbb{C}^{N\times N}$ be a sparse matrix.  If $A$ is
symmetric, the selected inversion
algorithm~\cite{TakahashiFaganChin1973,ErismanTinney1975,LinYangMezaEtAl2011,JacquelinLinYang2017}
and its
variants~\cite{CampbellDavis1995,LiAhmedKlimeckDarve2008,LiDarve2012,HetmaniukZhaoAnantram2013,AmestoyDuffLExcellentEtAl2012,AmestoyDuffLExcellentEtAl2015,PetersenLiStokbroEtAl2009,LinLuYingCarE2009,KuzminLuisierSchenk2013}
are efficient ways for computing certain selected elements of $A^{-1}$,
defined as $(A^{-1})_S:= \{(A^{-1})_{i,j}\vert \ \ \mbox{for} \ \ 1\le i,j\le N, \ \
\mbox{such that} \ \ A_{i,j}\ne 0 \}$. The algorithm actually computes more elements of $A^{-1}$ than $(A^{-1})_S$. The set of computed elements is a superset of $(A^{-1})_S$,
defined as $\{(A^{-1})_{i,j}\vert \ (L+U)_{i,j}\ne
0 \}$. Here, for simplicity, we have omitted the range of indices for
$i,j$. The $LU$ factorization of $A$ is given by $A=LU$, and the sparsity pattern of $U$ is the same as
that of $L^{T}$.   
Selected inversion algorithms
have already been used in a number of applications
such as density functional
theory~\cite{LinLuYingCarE2009,LinChenYangEtAl2013,LinGarciaHuhsEtAl2014}, quantum transport
theory~\cite{LiAhmedKlimeckDarve2008,LiDarve2012,HetmaniukZhaoAnantram2013,KuzminLuisierSchenk2013}, dynamical mean field theory
(DMFT)~\cite{KotliarSavrasovHauleEtAl2006}, Poisson-Boltzmann
equation~\cite{XuMaggs2013}, to name a few.  

In~\cite{ErismanTinney1975}, Erisman and Tinney demonstrated that a selected inversion procedure can be applied to non-symmetric matrices. In such a case, the selected inversion
algorithm computes $\{(A^{-1})_{i,j}\vert \ (L+U)_{j,i}\ne 0 \}$, and
therefore the definition of selected elements should be modified to
\begin{equation}
(A^{-1})_S:=\{(A^{-1})_{i,j}\vert A_{j,i}\ne 0 \}.
\label{eq.sel_entries}
\end{equation} 
Let us consider two extreme cases. 1) 
When $A$ is symmetric, the general definition of selected elements agree
with the previous definition. The same argument holds for
structurally symmetric matrices (i.e. $A_{i,j}\ne 0
\Leftrightarrow A_{j,i}\ne 0$). 2) When $A$ is an upper triangular or a
lower triangular matrix, the selected inversion algorithm only
computes the diagonal elements of $A^{-1}$. Indeed, these entries are
easy to compute since $(A^{-1})_{i,i}=(A_{i,i})^{-1}$, while
$\{(A^{-1})_{i,j}\vert \ A_{i,j}\ne 0 \}$ would include all the nonzero
entries of $A^{-1}$.

At first glance, it may seem restrictive that the selected inversion
algorithm for general matrices cannot even compute the entries of
$A^{-1}$ corresponding to the sparsity pattern of $A$.
Fortunately, this modified definition of selected elements is already
sufficient in a number of applications. One case is the computation of the
diagonal elements of $A^{-1}$. Another case is the computation of 
traces of the form 
$\Tr[B A^{-1}]=\sum_{ij} B_{i,j}(A^{-1})_{j,i}$, where the sparsity
pattern of $B\in\mathbb{C}^{N\times N}$ is contained in the sparsity
pattern of $A$, i.e. $\{(i,j)\vert B_{i,j}\ne 0\}\subset \{(i,j)\vert
A_{i,j}\ne 0\}$. This type of trace calculation appears in a number of contexts,
such as the computation of electron energy in density functional theory
calculations.  It is also a useful way to numerically
validate the identity $\Tr[A A^{-1}] = N$, which serves as a quick and useful
indicator of the accuracy of the computed selected elements of $A^{-1}$,
especially for large matrices of which the full inverse is too expensive
to compute.

Although the non-symmetric version of the selected inversion algorithm was proposed more than four decades ago, to
our knowledge, there is no efficient implementation of the selected inversion algorithm for general non-symmetric matrices, either sequential or parallel.  This paper fills this gap by
extending the \texttt{PSelInv} implementation reported
in~\cite{JacquelinLinYang2017} to non-symmetric matrices and  on
distributed memory parallel architecture. We remark that 
such a general treatment may be of
interest even for symmetric matrices, when additional static 
pivoting is performed to improve numerical stability~\cite{GolubVan1996,LiDemmel2003}. 
In such cases, the selected inversion algorithm needs to be applied to
the non-symmetric matrix $\wt{A}=PAQ$, where $P,Q$ are permutation matrices.

There are some notable differences between the implementation of
\pselinv for symmetric and non-symmetric matrices. First, a non-symmetric matrix only permits a $LU$ factorization, while both the $LU$ and
the $LDL^{T}$ factorization can be used for symmetric matrices.  
Second, for non-symmetric matrices, one can in principle perform a structural
symmetrization procedure by treating certain zero elements as nonzeros
and use the selected inversion algorithm for structurally symmetric
matrices.  However, such treatment is generally inefficient in terms of both the storage cost
and the computational cost. As an extreme case, structurally
symmetrizing an upper triangular matrix would mean that the matrix $A$ is
treated as a full dense square matrix. From this perspective, our parallel implementation is memory efficient, in the sense that no
symmetrization process is involved, and the selected elements of
$A^{-T}$ overwrites the sparse matrix $L+U$ \textit{in situ}. Here the
transpose corresponds to the definition of the selected
elements~\eqref{eq.sel_entries} and will be explained in detail later. 
Third, more complicated data communication
pattern is required to implement the parallel selected inversion
algorithm for non-symmetric matrices, and the selected elements of the inverse in the upper and lower triangular parts need to be treated separately. In~\cite{JacquelinLinYang2017} we explicitly take advantage of the
symmetry of the matrix to simplify some of the data communication. This
is no longer an option for non-symmetric matrices.
We develop a general point-to-point data communication strategy to 
efficiently handle collective data communication operations. This
general point-to-point data communication strategy allows us to use
a recently developed tree based asynchronous collective communication
method to improve load balancing when a large number of cores 
are used, as recently demonstrated for the symmetric
case of the \pselinv algorithm~\cite{JacquelinLinWichmannEtAl2016}.
Our numerical results indicate that the non-symmetric
version of \pselinv can be scalable to up to $6400$ cores depending on
the size and sparsity of the matrix.  Our implementation of \pselinv is publicly
available\footnote{\url{http://www.pexsi.org/}, distributed under the
BSD license}.

The rest of the paper is organized as follows.  We review the basic
idea of the selected inversion method for non-symmetric matrices in
section~\ref{sec:selinv}, and discuss various implementation issues for
the distributed memory parallel selected inversion algorithm for
non-symmetric matrices in
section~\ref{sec:parallelization}.  The numerical results with
applications to various matrices from including Harwell-Boeing Test Collection
\cite{HarwellBoeing}, the University of Florida Matrix
Collection\cite{FloridaMatrix}, and from density
functional theory in section~\ref{sec:numerical}, followed by
the conclusion and the future work discussion in section~\ref{sec:conclusion}.

Standard linear algebra notation is used for vectors and matrices
throughout the paper.  We use $A_{i,j}$ to denote the $(i,j)$-th entry
of the matrix $A$, and $f_{i}$ to denote the $i$-th entry of the vector
$f$. With slight abuse of notation, both a supernodal index and the set
of column indices associated with a supernode are denoted by uppercase
script letters such as $\IS,\JS,\KS$ \etc.  
%Furthermore, we use
%$A_{i,*}$ and $A_{*,j}$ to denote the $i$-th row and the $j$-th column
%of $A$, respectively. Similarly, $A_{\IS,*}$ and $A_{*,\JS}$ are used to
%denote the $\IS$-th block row and the $\JS$-th block column of $A$,
%respectively. 
$A_{\IS,\JS}^{-1}$ denotes the $(\IS,\JS)$-th block of the
matrix $A^{-1}$, i.e. $A_{\IS,\JS}^{-1}\equiv (A^{-1})_{\IS,\JS}$.  When
the block $A_{\IS,\JS}$ itself is invertible, its inverse is denoted by
$(A_{\IS,\JS})^{-1}$ to distinguish from $A_{\IS,\JS}^{-1}$. We also use
$A^{-T}_{\IS,\JS}$ to denote the $(\IS,\JS)$-th matrix block of the
transpose of the matrix $A^{-1}$.

\section{Selected inversion algorithm for non-symmetric matrices}\label{sec:selinv}

The standard approach for computing $A^{-1}$ is to first decompose
$A$ using the LU factorization 
\begin{equation}
	A = LU
	\label{eqn:LU}
\end{equation}
where $L$ is a unit lower triangular matrix and $U$ is an upper
triangular matrix.  In order to stabilize the computation, matrix
reordering and row pivoting (or partial pivoting)~\cite{GolubVan1996} are usually applied to the
matrix of A, and the general form of the $LU$ factorization can be
given as
\begin{equation}
  PAQ = \wt{A} = LU,
	\label{eqn:permuteLU}
\end{equation}
where $P$ and $Q$ are two permutation matrices.  Care must be taken when 
non-symmetric row and column permutations are used, i.e. $P\ne Q^{T}$.
To simplify the discussion for now, we use Eq.~\eqref{eqn:LU} and assume
$A$ has already been permuted.

The selected inversion algorithm can be \textit{heuristically} understood as follows. We
first partition the matrix $A$ into $2\times 2$
blocks of the form
\begin{equation}
	A = \begin{pmatrix}
		A_{1,1} & A_{1,2}\\
		A_{2,1} & A_{2,2}
	\end{pmatrix},
	\label{}
\end{equation}
where $A_{1,1}$ is a scalar of size $1\times 1$.  We can
write $A_{1,1}$ as a product of two scalars $L_{1,1}$ and $U_{1,1}$.
In particular, we can pick $L_{1,1}=1$ and $U_{1,1}=A_{1,1}$. Then 
\begin{equation}
	A = \begin{pmatrix}
		L_{1,1} & 0\\
		L_{2,1} & I 
	\end{pmatrix}
	\begin{pmatrix}
		U_{1,1} & U_{1,2}\\
		0      & S_{2,2}
	\end{pmatrix}
	\label{eqn:LU2by2}
\end{equation}
where
\begin{equation}
	L_{2,1}=A_{2,1} (U_{1,1})^{-1}, \quad U_{1,2} = (L_{1,1})^{-1}	A_{1,2},
	\label{}
\end{equation}
and
\begin{equation}
	S_{2,2} = A_{2,2} - L_{2,1} U_{1,2}	
	\label{}
\end{equation}
is the Schur complement.

Using the decomposition given by 
Eq.~\eqref{eqn:LU2by2}, we can express $A^{-1}$ as
\begin{equation}
	A^{-1} = \begin{pmatrix}
		(U_{1,1})^{-1} (L_{1,1})^{-1} + (U_{1,1})^{-1} U_{1,2} S^{-1}_{2,2}
		L_{2,1} (L_{1,1})^{-1} & - (U_{1,1})^{-1} U_{1,2} S^{-1}_{2,2} \\
		-S^{-1}_{2,2} L_{2,1} (L_{1,1})^{-1} & S_{2,2}^{-1}
	\end{pmatrix}.
	\label{eqn:Ainv2by2}
\end{equation}
%Since $S_{2,2}$ is the same as $S$ here, without ambiguity $S_{2,2}^{-1}\equiv (S^{-1})_{2,2}$ can be used, and $A^{-1}_{2,2} = S^{-1}_{2,2}$.
With slight abuse of notation, define $\CS_{L} :=  \{i\vert L_{i,1}\ne 0 \}$ and  $\CS_{U} :=  \{j\vert
U_{1,j}\ne 0\}$.
Here $L_{i,1}$ is the $i$-th component of the column vector
$(L_{1,1}, L_{2,1})^T$ as in Eq.~\eqref{eqn:LU2by2}, and $U_{1,j}$ is
the $j$-th component of the row vector $(U_{1,1}, U_{1,2})$. The sets $\CS_{L}$ and $\CS_{U}$ are defined purely in terms of the nonzero structures of $L$ and $U$, i.e.,
$L_{i,1}$ and $U_{1,j}$ treated as nonzeros even if their numerical values are coincidentally $0$.  For non-symmetric matrices, $\CS_{L}$ and $\CS_{U}$ may not be the same. 

We assume $S_{2,2}^{-1}$ has already been computed. 
From Eq.~\eqref{eqn:Ainv2by2} it can be readily observed that, if $L$ and $U$ are sparse, the $(1,1)$ entry of $A^{-1}$ can be computed from the nonzero elements of $L_{2,1}$ and $U_{1,2}$ together with the corresponding selected entries of $S_{2,2}^{-1}$. Because $A^{-1}_{2,2} = S^{-1}_{2,2}$, the selected entries of $S_{2,2}^{-1}$ belong to a subset of 
\begin{equation}
    \left\{ A^{-1}_{i,j} \vert  i\in \CS_{U}, j\in
    \CS_{L}\right\}.
	\label{eqn:selectentry2x2}
\end{equation}
which also include  
$\{A^{-1}_{1,j}  \vert j\in \CS_{L}\}$ and
$\{A_{i,1}^{-1} \vert i \in \CS_{U}\}$. The latter can be computed from the same selected elements of $S_{2,2}^{-1}$, $L_{2,1}$ and $U_{1,2}$. Repeating the procedure above recursively for $S_{2,2}$, we can see how the selected elements of $A_{i,k}^{-1}$ and $A_{k,j}^{-1}$ that are required to compute the selected elements of $A^{-1}$ in the rows and columns preceding $k$ can be computed from selected elements of the trailing $(n-k)\times (n-k)$ block of $A^{-1}$.  
This argument can be stated more precisely in Theorem~\ref{thm:selinv}.

\begin{theorem}[Erisman and Tinney~\cite{ErismanTinney1975}]
\label{thm:selinv}
For a matrix $A\in\mathbb{C}^{N\times N}$, let $A=LU$ be its
$LU$ factorization, and $L,U$ are invertible matrices. For any
$1\le k < N$, define
\begin{align}
    \CS_{L} =  \{i\vert L_{i,k}\ne 0\}, \quad \CS_{U} =  \{j\vert U_{k,j} \ne 0\}.
    \label{}
\end{align}
Then all entries $\{A^{-1}_{i,k} \vert i\in \CS_{U}\}$,  
$\{A^{-1}_{k,j} \vert j\in \CS_{L}\}$, and $A^{-1}_{k,k}$ can be
computed using only $\{L_{j,k}| j\in \CS_{L}\}$, $\{U_{k,i}| i\in \CS_U\}$ and $\{A^{-1}_{i,j} \vert (L+U)_{j,i} \ne 0,
i, j\ge k\}$. %\CY{do we need to distinguish $i$, $j$, $p$, $q$}
\end{theorem}
\begin{proof}
  First consider $\{A^{-1}_{i,k} \vert i\in
  \CS_{U}\}$. Similar to Eq.~\eqref{eqn:Ainv2by2} we can derive
  \begin{equation}
    A^{-1}_{i,k} = -\sum_{j=k+1}^{N} A^{-1}_{i,j} L_{j,k}
    (L_{k,k})^{-1}, \quad i\in \CS_{U}.
    \label{eqn:selinv_1}
  \end{equation}
  If $L_{j,k}\ne 0$, then $A^{-1}_{i,j}$ is needed in the sum. Since we
  are only interested in computing $A^{-1}_{i,k}$ for $i\in
  \CS_{U}$, the $i$ and $j$ indices are constrained to satisfy the conditions $L_{j,k}\ne 0$ and $U_{k,i}\ne 0$. This constraint implies $(L+U)_{j,i}\ne 0$ because the nonzero fill-in pattern of the trailing blocks of $L$ and $U$ are determined by the nonzero patterns of the $k$th column of $L$ and the $k$th row of $U$ respectively. 
 A similar argument can be made for $\{A^{-1}_{k,j} \vert
  j\in \CS_{L}\}$. Finally for the diagonal entry, we have
  \begin{equation}
    A^{-1}_{k,k} = (U_{k,k})^{-1} (L_{k,k})^{-1} - 
    \sum_{i=k+1}^{N} (U_{k,k})^{-1} U_{k,i} A^{-1}_{i,k},
    \label{eqn:selinv_2}
  \end{equation}
  which can be readily computed given $\{A^{-1}_{i,k} \vert i\in
  \CS_{U}\}$ is available. 
\end{proof}

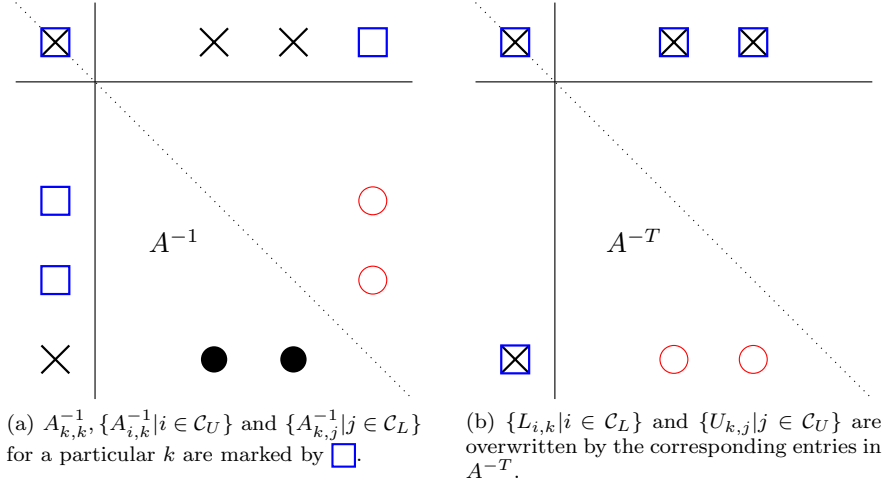
\begin{figure}[htbp]
\centering
\subfloat[$A^{-1}_{k,k},\{A^{-1}_{i,k}| i\in \CS_{U}\}$ and $\{A_{k,j}^{-1}|j \in \CS_{L}\}$ for a particular $k$ are marked by \linvsquare.]{
\begin{adjustbox}{width=.35\linewidth}
\begin{tikzpicture}
\begin{scope}[yscale=-1]
\draw (0,0) node[cross,rotate=0] {};
\draw (0.5,-0.5) -- (0.5,4.5);
\draw (-0.5,0.5) -- (4.5,0.5);
\draw[dotted] (-0.5,-0.5) -- (4.5,4.5);

\draw (0,4) node[cross,rotate=0] {};
%\draw (0,5) node[cross,rotate=0] {};
%\draw (0,6) node[cross,rotate=0] {};

\draw (2,0) node[cross,rotate=0] {};
\draw (3,0) node[cross,rotate=0] {};
%\draw (5,0) node[cross,rotate=0] {};

\draw (3,4) node[circle,minimum size=.2em,fill=black] {};
\draw (2,4) node[circle,minimum size=.2em,fill=black] {};

\draw (4,2) node[circle,minimum size=1em,rotate=0,draw=red] {};
%\draw (5,2) node[circle,minimum size=1em,rotate=0,draw=red] {};
%\draw (6,2) node[circle,minimum size=1em,rotate=0,draw=red] {};
\draw (4,3) node[circle,minimum size=1em,rotate=0,draw=red] {};
%\draw (5,3) node[circle,minimum size=1em,rotate=0,draw=red] {};
%\draw (6,3) node[circle,minimum size=1em,rotate=0,draw=red] {};
%\draw (4,5) node[circle,minimum size=1em,rotate=0,draw=red] {};
%\draw (5,5) node[circle,minimum size=1em,rotate=0,draw=red] {};
%\draw (6,5) node[circle,minimum size=1em,rotate=0,draw=red] {};

\draw (0,0) node[rectangle, minimum size = 1em,rotate=0,draw=blue,thick] {};
\draw (4,0) node[rectangle, minimum size = 1em,rotate=0,draw=blue,thick] {};
%\draw (5,0) node[rectangle, minimum size = 1em,rotate=0,draw=blue,thick] {};
%\draw (6,0) node[rectangle, minimum size = 1em,rotate=0,draw=blue,thick] {};
\draw (0,2) node[rectangle, minimum size = 1em,rotate=0,draw=blue,thick] {};
\draw (0,3) node[rectangle, minimum size = 1em,rotate=0,draw=blue,thick] {};
%\draw (0,5) node[rectangle, minimum size = 1em,rotate=0,draw=blue,thick] {};

\node at( 1.5, 2.5) {$A^{-1}$};

\end{scope}
\end{tikzpicture}
\end{adjustbox}
}
~~~~
\subfloat[$\{L_{i,k}| i\in \CS_{L}\}$ and $\{U_{k,j}| j\in \CS_{U}\}$ are overwritten by the corresponding entries in $A^{-T}$.]{
\begin{adjustbox}{width=.35\linewidth}
\begin{tikzpicture}
\begin{scope}[yscale=-1]
\draw (0,0) node[cross,rotate=0] {};
\draw (0.5,-0.5) -- (0.5,4.5);
\draw (-0.5,0.5) -- (4.5,0.5);
\draw[dotted] (-0.5,-0.5) -- (4.5,4.5);

\draw (0,4) node[cross,rotate=0] {};
%\draw (0,5) node[cross,rotate=0] {};
%\draw (0,6) node[cross,rotate=0] {};

\draw (2,0) node[cross,rotate=0] {};
\draw (3,0) node[cross,rotate=0] {};
%\draw (5,0) node[cross,rotate=0] {};

%\draw (4,2) node[cross,rotate=0,draw=black,dotted] {};
%\draw (5,2) node[cross,rotate=0,draw=black,dotted] {};
%\draw (6,2) node[cross,rotate=0,draw=black,dotted] {};
%\draw (4,3) node[cross,rotate=0,draw=black,dotted] {};
%\draw (5,3) node[cross,rotate=0,draw=black,dotted] {};
%\draw (6,3) node[cross,rotate=0,draw=black,dotted] {};
%\draw (4,5) node[cross,rotate=0,draw=black,dotted] {};
%\draw (5,5) node[cross,rotate=0,draw=black,dotted] {};
%\draw (6,5) node[cross,rotate=0,draw=black,dotted] {};

\draw (2,4) node[circle,minimum size=1em,rotate=0,draw=red] {};
%\draw (2,5) node[circle,minimum size=1em,rotate=0,draw=red] {};
%\draw (2,6) node[circle,minimum size=1em,rotate=0,draw=red] {};
\draw (3,4) node[circle,minimum size=1em,rotate=0,draw=red] {};
%\draw (3,5) node[circle,minimum size=1em,rotate=0,draw=red] {};
%\draw (3,6) node[circle,minimum size=1em,rotate=0,draw=red] {};
%\draw (5,4) node[circle,minimum size=1em,rotate=0,draw=red] {};
%\draw (5,5) node[circle,minimum size=1em,rotate=0,draw=red] {};
%\draw (5,6) node[circle,minimum size=1em,rotate=0,draw=red] {};

\draw (0,0) node[rectangle, minimum size=1em, draw=blue,thick] {};
\draw (0,4) node[rectangle, minimum size=1em, draw=blue,thick] {};
%\draw (0,5) node[rectangle, minimum size=1em, draw=blue,thick] {};
%\draw (0,6) node[rectangle, minimum size=1em, draw=blue,thick] {};
\draw (2,0) node[rectangle, minimum size=1em, draw=blue,thick] {};
\draw (3,0) node[rectangle, minimum size=1em, draw=blue,thick] {};
%\draw (5,0) node[rectangle, minimum size=1em, draw=blue,thick] {};

\node at( 1.5, 2.5) {$A^{-T}$};

\end{scope}
\end{tikzpicture}
\end{adjustbox}
}
\caption{(a)   
$\{L_{i,k}|i\in \CS_{L}\}$ and $\{U_{k,j}| j\in \CS_{U}\}$ are marked by
\lucross. The nonzero fills introduced by the $k$th column of $L$ and $k$th row of $U$ are represented by \ainvfill. 
(b) $A^{-T}$ in the selected inversion algorithm can directly overwrite
the $L,U$ factors.
%\LL{Red and black anticrosses are not distinguishable on
%black-and-white. How about changing the red anticross to circles? The
%blue square should be on the diagonal as well? Also 
%in the picture maybe indicate the fill-in entries of L and U corresponding to
%the black crosses. }
\label{fig:selinv_entries}}
\end{figure}

Fig.~\ref{fig:selinv_entries} (a) illustrates one step of the selected inversion
procedure for a general matrix. For example, according to
Eq.~\eqref{eqn:selinv_1} and~\eqref{eqn:selinv_2}, computing the $(k,k$)th element of $A^{-1}$ shown at the upper left corner of the figure requires previously computed element of $A^{-1}$ marked by red circles. To compute $A_{k,k}^{-1}$ we effectively have to compute the selected element of $A^{-1}$ marked by the blue squares. By pretending that we are computing the selected elements of $A^{-T}$ instead, we can overwrite the corresponding elements of $U$ and $L$ as shown in Fig.~\ref{fig:selinv_entries} (b).  
Theorem~\ref{thm:selinv} directly indicates that any element of $A^{-1}$
corresponding to the sparsity pattern of $(L+U)^{T}$ can be evaluated
using $L,U$ and other elements of $A$ in this subset of entries. In
particular, the selected elements $\{A^{-1}_{i,j}\vert A_{j,i}\ne 0
\}$ can be evaluated efficiently. 

So far we have not explicitly taken into account row and column
permutation. Theorem~\ref{thm:selinvperm} demonstrates that the same
result holds when permutation is involved.

\begin{theorem}
\label{thm:selinvperm}
For $A\in\mathbb{C}^{N\times N}$, let $PAQ=\wt{A}=LU$ be its
$LU$ factorization. Here $L,U$ are invertible matrices, and $P,Q$ are
permutation matrices. Then $\{A^{-1}_{i,j}\vert A_{j,i}\ne 0
\}$ can be evaluated using $L,U$ and $\{\wt{A}^{-1}_{i,j} \vert (L+U)_{j,i}
\ne 0\}$.
\end{theorem}
\begin{proof}
  Since $P,Q$ are permutation matrices, $PP^{T}=QQ^{T}=I$, and we have
  the identity
  \begin{equation}
    A^{T}=Q\wt{A}^{T} P, \quad A^{-1}=Q\wt{A}^{-1}P.
    \label{eqn:unpermute}
  \end{equation}
  Since the entries $\{\wt{A}^{-1}_{i,j} \vert \wt{A}_{j,i} \ne 0\}
  \subset \{\wt{A}^{-1}_{i,j} \vert (L+U)_{j,i} \ne 0\}$ have
  been computed, undo the permutation of $\wt{A}^{-1}$ and we obtain
  $\{A^{-1}_{i,j} \vert A_{j,i}\ne 0\}$, which are the required selected elements of $A^{-1}$.
\end{proof}

In practice, a column-based sparse factorization and selected
inversion algorithm may not be efficient due to the lack of level 3 BLAS 
operations.  For a sparse matrix $A$, 
the columns of $A$ and the $L$ factor can be
partitioned into supernodes. A supernode is a maximal set of contiguous
columns $\JS=\{j,j+1,\ldots,j+s\}$ of the $L$ factor that have the
same nonzero structure below the $(j+s)$-th row, and the lower
triangular part of $L_{\JS,\JS}$ is dense. However, this strict 
definition can produce supernodes that are either too large or too small, 
leading to memory usage, load balancing and efficiency issues.
Therefore, in our work, we relax this definition to
limit the maximal number of columns in a supernode (i.e. sets are not
necessarily maximal).  The relaxation also allows a supernode to include
columns for which nonzero patterns are nearly identical to enhance the
efficiency~\cite{AshcraftGrimes1989}, and this approach is also used in
\superlu~\cite{LiDemmel2003}.  We assume the same supernode partitioning is usually
applied to the row partition as well, even though the nonzero pattern of the
$L$ and $U$ can be different from each other.  The total
number of supernodes is denoted by $\mathcal{N}$.  
Using the notation of supernodes (e.g. $1$ means the first supernode
instead of the first column index), $L_{1,1}$ is no longer a scalar
$1$ or an identity matrix, but a lower triangular matrix.
To simplify the notation of the selected inversion algorithm, in
Eq.~\eqref{eqn:Ainv2by2} we can define
the normalized $LU$ factors as 
\begin{equation}
	\hat{L}_{1,1} = L_{1,1}, \quad \hat{U}_{1,1}=U_{1,1},\quad
	\hat{L}_{2,1} = L_{2,1}(L_{1,1})^{-1}, \quad \hat{U}_{1,2} = (U_{1,1})^{-1}
	U_{1,2}.
	\label{}
\end{equation}
This definition can be directly generalized for other columns and for
the case when supernodes are used.
Furthermore, from an implementation perspective, the definition of selected
elements indicates that it is most natural to formulate the selected
inversion algorithm to compute $A^{-T}$, so that $A^{-T}$ can directly
overwrite the $L,U$ factors (see Fig.~\ref{fig:selinv_entries} (b)). A pseudo-code for the selected inversion
algorithm for non-symmetric matrices is given in
Alg.~\ref{alg:selinvlu},
which can readily be used as a sequential
implementation of the selected inversion algorithm.
Note that in
step \ref{alg1.step4}, 
the diagonal entry can be equivalently computed using the formula
$\wt{A}^{-T}_{\KS,\KS} \gets (L_{\KS,\KS})^{-T} (U_{\KS,\KS})^{-T} 
    - (\hat{L}_{\CS_{L},\KS})^{T}\wt{A}^{-T}_{\CS_{L},\KS}$.
We also note that the normalized  factors $\hat{L},\hat{U}$ can overwrite the
$L,U$ factors, and the intermediate matrix $\wt{A}^{-T}$ can overwrite
the normalized factors whenever the computation for a given supernode
$\KS$ is finished. However, we keep these matrices with distinct
notations in Alg.~\ref{alg:selinvlu} for clarity. 

\begin{algorithm}
  \DontPrintSemicolon
  \caption{Selected inversion algorithm for a general sparse matrix $A$.}
  \label{alg:selinvlu}

  \KwIn{\begin{tabular}{l} 
    (1) \begin{minipage}[t]{4.0in} Permutation matrices $P,Q$.
    \end{minipage}\\
    (2) \begin{minipage}[t]{4.0in} The supernodal
      partition $\{1,2,...,\mathcal{N}\}$. \end{minipage}\\
    (3) \begin{minipage}[t]{4.0in} A supernodal sparse $LU$ factorization
      $PAQ=\wt{A}=LU$. \end{minipage}
    \end{tabular}
      }

   \KwOut{\begin{minipage}[t]{4.0in} $\{A^{-1}_{\IS,\JS} 
       \vert A_{\JS,\IS} \mbox{ is a nonzero block},
       \IS,\JS=1,\ldots,\mathcal{N}\}$. \end{minipage} } 

	\For{$\KS = \mathcal{N}, \mathcal{N}-1, ..., 1$}{
    Find the collection of indices\;
    $\qquad \CS_{L}=\{\IS\vert\IS>\KS,L_{\IS,\KS}\mbox{ is a
	nonzero block}\}$\;
    $\qquad \CS_{U}=\{\JS\vert\JS>\KS, U_{\KS,\JS}\mbox{ is a
	nonzero block}\}$\;
    \lnl{alg1.step2} $\hat{L}_{\CS_{L},\KS}\gets L_{\CS_{L},\KS} (L_{\KS,\KS})^{-1},
    \hat{U}_{\KS,\CS_{U}}\gets (U_{\KS,\KS})^{-1} U_{\KS,\CS_{U}}$\; 
  }

	\For{$\KS = \mathcal{N}, \mathcal{N}-1, ..., 1$}{
    Find the collection of indices\;
    $\qquad \CS_{L}=\{\IS\vert\IS>\KS,L_{\IS,\KS}\mbox{ is a
	nonzero block}\}$\;
    $\qquad \CS_{U}=\{\JS\vert\JS>\KS, U_{\KS,\JS}\mbox{ is a
	nonzero block}\}$\;
  \lnl{alg1.step3} Calculate $\wt{A}^{-T}_{\KS,\CS_{U}} \gets
  -(\hat{L}_{\CS_{L},\KS})^{T} \wt{A}^{-T}_{\CS_{L},\CS_{U}}$\; 
  \lnl{alg1.step4} Calculate $\wt{A}^{-T}_{\KS,\KS} \gets 
    (L_{\KS,\KS})^{-T} (U_{\KS,\KS})^{-T}
    - \wt{A}^{-T}_{\KS,\CS_{U}}(\hat{U}_{\KS,\CS_{U}})^{T}$\;
%    - (\hat{L}_{\CS_{L},\KS})^{T}\wt{A}^{-T}_{\CS_{L},\KS}$\; 
  \lnl{alg1.step5} Calculate $\wt{A}^{-T}_{\CS_{L},\KS} \gets -
    \wt{A}^{-T}_{\CS_{L},\CS_{U}}(\hat{U}_{\KS,\CS_{U}})^{T}$\;
  }
  \lnl{alg1.step6}Extract the matrix blocks $\{\wt{A}^{-T}_{\JS,\IS} 
  \vert \wt{A}_{\JS,\IS} \mbox{ is a nonzero block}\}$, undo the
  permutation and apply matrix transpose to obtain $\{A^{-1}_{\IS,\JS} 
  \vert A_{\JS,\IS} \mbox{ is a nonzero block},
  \IS,\JS=1,\ldots,\mathcal{N}\}$
\end{algorithm}

\section{Distributed memory parallel selected inversion algorithm for
non-symmetric matrices}\label{sec:parallelization}

%\LL{Main points in this section
%\begin{itemize}
%  %\item Data structure layout using an example of an asymmetric matrix.
%  %\item Computational procedure for updating one supernode. Emphasize
%  %  the algorithmic difference: maybe briefly summarizing the procedure
%  %  and cite the previous paper for details, and showing Fig. 4 with the
%  %  arrows of reverse directions to illustrate the update of U? Update
%  %  the parallel selected inversion algorithm.
%  \item Discuss the partition of ``small tasks'', which removes all the
%    MPI barriers in the code, and prepares for dynamic scheduling? The
%    difference in terms of communication pattern also reduces the memory
%    footprint.
%\end{itemize}
%}

%In this section we present the distributed memory of the general non-symmetric
%\pselinv method. 

In this section, we present the \pselinv method for general non-symmetric
matrices on distributed memory parallel architecture.
The selected inversion algorithm
described in Alg.~\ref{alg:selinvlu} requires a sparse $LU$
factorization of the permuted matrix $\wt{A}=PAQ$ to be computed first. 
We compute the $LU$ decomposition using the \superlu
software package~\cite{LiDemmel2003}, which has been shown to be scalable to
a large number of processors on distributed memory parallel machines.
\superlu allows 
the sparse $L$ and $U$ factors to be accessed through relatively simple data structures.
However, it should be noted that the ideas developed in this section 
can be combined with other sparse matrix solvers such as \mumps~\cite{mumps} or
\pardiso~\cite{pardiso} too, provided that the factors are available.
%The implementation of the non-symmetric \pselinv method is optimal in terms
%of computation. It is also optimal in terms of memory consumption, contrarily
%to our symmetric implementation.\LL{mention this later with some more
%clarity}
 
As discussed at the end of section~\ref{sec:selinv}, in order to achieve
a memory efficient implementation, we work with the transposed matrix
inverse $\wt{A}^{-T}$, which can directly overwrite the $LU$ factors. To
simplify the notation, in this section we do not distinguish $A$ and the
permuted matrix $\wt{A}$.
We use the same 2D
block cyclic distribution scheme employed in \superlu to partition and
distribute both the $L,U$ factors and the selected elements of $A^{-T}$
to be computed. We will review the main features of this type of
distribution in section~\ref{subsec:datadist}.  In the 2D block cyclic
distribution scheme, each supernode $\mathcal{K}$ is assigned to and
partitioned among a subset of processors. However, computing the
selected elements of $A^{-T}$ associated with the supernode $\mathcal{K}$
requires retrieving some previously computed selected elements of
$A^{-T}$ that belong to ancestors of $\mathcal{K}$ in the elimination
tree. These selected elements may reside on other processors. As a
result, communication is required to transfer data among different
processors to complete steps \ref{alg1.step3} to
\ref{alg1.step5} of Alg.~\ref{alg:selinvlu} in
each iteration.  We will discuss how this is done in
section~\ref{subsec:block}.
Furthermore, in order to achieve scalable performance on thousands of
cores, it
is important to overlap communication with computation using asynchronous 
point-to-point MPI functions. In the \pselinv method, most of these
communication operations
%even though some of these communication 
are collective in nature (e.g., broadcast and reduce) within 
communication subgroups. The sizes of the communication groups can vary
widely for operations associated with different supernodes. We will describe how such
collective communication operations can be efficiently performed
asynchronously in section~\ref{subsec:commun}.

%\MJ{Should we talk about the limitation on the number of communicators
% to clearly state that we can only have collectives on row/columns
% but not on sparse communicators? Also, provide more info on the async communication trees ?}

% LL: The following paragraph is commented out since it is out-of-place
%
%Similar to our symmetric implementation, we use two levels of parallelism:
%a fine grain level of parallelism in computing $A^{-T}$ for each supernode,
%and a coarse grain level of parallelism exploiting the concurrency available
%in the elimination tree. 
%This amounts to executing different iterates of the {\tt for} loop in
%Alg.~\ref{alg:selinvlu} in parallel.  Although the elimination
%tree may exhibit many independent tasks associated with supernodes that
%belong to different branches of the elimination tree, the 2D block cyclic
%distribution of $L$, $U$ and $A^{-T}$ may prevent these
%tasks from being performed completely simultaneously on different
%processors. We introduced a preliminary strategy for improving the parallel
%efficiency using elimination tree in~\MJ{citation}. We use the same strategy
%for our non-symmetric implementation.

\subsection{Distributed data layout and structure} \label{subsec:datadist}

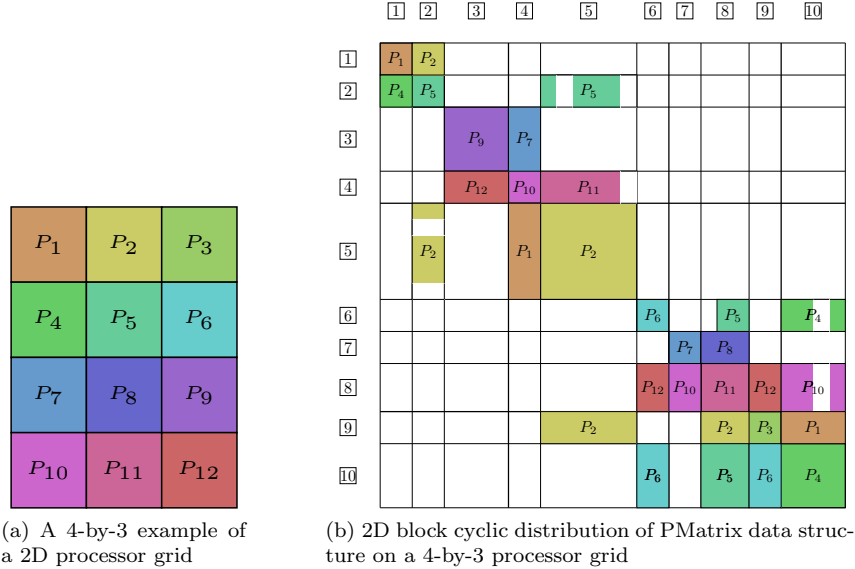
\begin{figure}[htbp]
\centering

\subfloat[A 4-by-3 example of a 2D processor grid]{
\label{fig:pmatrix_grid}
\begin{adjustbox}{width=.20\linewidth}
%\scalefont{0.5}
\begin{tikzpicture}[every node/.style={font=\tiny}]

    \pgfmathtruncatemacro{\Pr}{4};
    \pgfmathtruncatemacro{\Pc}{3};

\mapblock{1}{1}{\Pr}{\Pc}
\colorproc{\curproc}{\Pr}{\Pc}
\blk{\Proc{\curproc}}{2}{2}{1}{1}{,fill=proccolor};
\mapblock{1}{2}{\Pr}{\Pc}
\colorproc{\curproc}{\Pr}{\Pc}
\blk{\Proc{\curproc}}{2}{2}{1}{3}{,fill=proccolor};
\mapblock{1}{3}{\Pr}{\Pc}
\colorproc{\curproc}{\Pr}{\Pc}
\blk{\Proc{\curproc}}{2}{2}{1}{5}{,fill=proccolor};
\mapblock{1}{4}{\Pr}{\Pc}
\colorproc{\curproc}{\Pr}{\Pc}
\blk{\Proc{\curproc}}{2}{2}{1}{7}{,fill=proccolor};

\mapblock{2}{1}{\Pr}{\Pc}
\colorproc{\curproc}{\Pr}{\Pc}
\blk{\Proc{\curproc}}{2}{2}{3}{1}{,fill=proccolor};
\mapblock{2}{2}{\Pr}{\Pc}
\colorproc{\curproc}{\Pr}{\Pc}
\blk{\Proc{\curproc}}{2}{2}{3}{3}{,fill=proccolor};
\mapblock{2}{3}{\Pr}{\Pc}
\colorproc{\curproc}{\Pr}{\Pc}
\blk{\Proc{\curproc}}{2}{2}{3}{5}{,fill=proccolor};
\mapblock{2}{4}{\Pr}{\Pc}
\colorproc{\curproc}{\Pr}{\Pc}
\blk{\Proc{\curproc}}{2}{2}{3}{7}{,fill=proccolor};

\mapblock{3}{1}{\Pr}{\Pc}
\colorproc{\curproc}{\Pr}{\Pc}
\blk{\Proc{\curproc}}{2}{2}{5}{1}{,fill=proccolor};
\mapblock{3}{2}{\Pr}{\Pc}
\colorproc{\curproc}{\Pr}{\Pc}
\blk{\Proc{\curproc}}{2}{2}{5}{3}{,fill=proccolor};
\mapblock{3}{3}{\Pr}{\Pc}
\colorproc{\curproc}{\Pr}{\Pc}
\blk{\Proc{\curproc}}{2}{2}{5}{5}{,fill=proccolor};
\mapblock{3}{4}{\Pr}{\Pc}
\colorproc{\curproc}{\Pr}{\Pc}
\blk{\Proc{\curproc}}{2}{2}{5}{7}{,fill=proccolor};

\end{tikzpicture}
\end{adjustbox}
}
~~~~~~~~
\subfloat[2D block cyclic distribution of \pmat data structure on a
4-by-3 processor grid]{
\label{fig:pmatrix_layout}
\begin{adjustbox}{width=.45\linewidth}
\begin{tikzpicture}

    \pgfmathtruncatemacro{\Pr}{4};
    \pgfmathtruncatemacro{\Pc}{3};

\node at (-2 em, -1em) {\supernode{1}};
\node at (-2 em, -3em) {\supernode{2}};
\node at (-2 em, -6em) {\supernode{3}};
\node at (-2 em, -9em) {\supernode{4}};
\node at (-2 em, -13em) {\supernode{5}};
\node at (-2 em, -17em) {\supernode{6}};
\node at (-2 em, -19em) {\supernode{7}};
\node at (-2 em, -21.5em) {\supernode{8}};
\node at (-2 em, -24em) {\supernode{9}};
\node at (-2 em, -27em) {\supernode{10}};

\node at (1em ,2 em) {\supernode{1}};
\node at (3em ,2 em) {\supernode{2}};
\node at (6em ,2 em) {\supernode{3}};
\node at (9em ,2 em) {\supernode{4}};
\node at (13em,2 em) {\supernode{5}};
\node at (17em,2 em) {\supernode{6}};
\node at (19em,2 em) {\supernode{7}};
\node at (21.5em,2 em) {\supernode{8}};
\node at (24em,2 em) {\supernode{9}};
\node at (27em,2 em) {\supernode{10}};
%D
\mdblk{}{1}{2}{1}{\Pr}{\Pc};
\mdblk{}{2}{2}{3}{\Pr}{\Pc};
\mdblk{}{3}{4}{6}{\Pr}{\Pc};
\mdblk{}{4}{2}{9}{\Pr}{\Pc};
\mdblk{}{5}{6}{13}{\Pr}{\Pc};
\mdblk{}{6}{2}{17}{\Pr}{\Pc};
\mdblk{}{7}{2}{19}{\Pr}{\Pc};
\mdblk{}{8}{3}{21.5}{\Pr}{\Pc};
\mdblk{}{9}{2}{24}{\Pr}{\Pc};
\mdblk{}{10}{4}{27}{\Pr}{\Pc};

\draw (0em,0em) --   (29em ,0em);
\draw (0em,-2em) --  (29em ,-2em);
\draw (0em,-4em) --  (29em ,-4em);
\draw (0em,-8em) --  (29em ,-8em);
\draw[thick] (0em,-10em) -- (29em ,-10em);
\draw (0em,-16em) -- (29em ,-16em);
\draw (0em,-18em) -- (29em ,-18em);
\draw (0em,-20em) -- (29em ,-20em);
\draw[thick] (0em,-23em) -- (29em ,-23em);
\draw (0em,-25em) -- (29em ,-25em);
\draw (0em,-29em) -- (29em ,-29em);

\draw (0em,0em) --  (0em ,-29em);
\draw (2em ,0em) -- (2em ,-29em);
\draw (4em ,0em) -- (4em ,-29em);
\draw[thick] (8em ,0em) -- (8em ,-29em);
\draw (10em,0em) -- (10em,-29em);
\draw (16em,0em) -- (16em,-29em);
\draw[thick] (18em,0em) -- (18em,-29em);
\draw (20em,0em) -- (20em,-29em);
\draw (23em,0em) -- (23em,-29em);
\draw[thick] (25em,0em) -- (25em,-29em);
\draw (29em,0em) -- (29em,-29em);

%L
%1
\mlublk{}{1}{2}{2}{2}{1}{3}{\Pr}{\Pc};
%2
\mlublk{}{2}{5}{2}{6}{3}{13}{\Pr}{\Pc};

\elublkFill{}{2}{1}{3}{11.5};
\elublkFill{}{2}{1}{3}{15.5};
\draw[draw=black] ($(15.5em,-3em) + 0.5*(1em,2em)$) -- ($(15.5em,-3em) + 0.5*(1em,-2em)$);
\draw[draw=black] ($(3em,-15.5em) + 0.5*(2em,-1em)$) -- ($(3em,-15.5em) + 0.5*(-2em,-1em)$);

%3
\mlublk{}{3}{4}{4}{2}{6}{9}{\Pr}{\Pc};
%4
\mlublk{}{4}{5}{2}{6}{9}{13}{\Pr}{\Pc};

\eublkFill{}{2}{1}{9}{15.5};

%5
\mlblk{}{5}{9}{6}{2}{13}{24}{\Pr}{\Pc};

%6
\mlblk{}{6}{8}{2}{3}{17}{21.5}{\Pr}{\Pc};
\mublk{}{6}{8}{2}{2}{17}{22}{\Pr}{\Pc};
\mlublk{}{6}{10}{2}{4}{17}{27}{\Pr}{\Pc};

\eublkFill{}{2}{1}{17}{27.5};
\node at (17em,-27em) {\Proc{6}};
\node at (27em,-17em) {\Proc{4}};

%7
\mlublk{}{7}{8}{2}{3}{19}{21.5}{\Pr}{\Pc};

%8
\mlublk{}{8}{9}{3}{2}{21.5}{24}{\Pr}{\Pc};
\mlublk{+}{8}{10}{3}{4}{21.5}{27}{\Pr}{\Pc};

\eublkFill{}{3}{1}{21.5}{27.5};
\node at (21.5em,-27em) {\Proc{5}};
\node at (27em,-21.5em) {\Proc{10}};
%9
\mlublk{}{9}{10}{2}{4}{24}{27}{\Pr}{\Pc};

\end{tikzpicture}
\end{adjustbox}
}

\caption{Data layout of the non-symmetric \pmat data structure used by \pselinv.}
\end{figure}

As discussed in Section~\ref{sec:selinv}, the columns of 
$A$, $L$ and $U$ are partitioned into supernodes.
Different supernodes may have different sizes.  
The same partition is 
applied to the rows of these matrices to create a 2D block partition of these 
matrices.  The submatrix blocks are mapped to processors that are 
arranged in a virtual 2D grid of dimension $\mathrm{Pr} \times
\mathrm{Pc}$ in a cyclic fashion as follows:  The $(\IS,\JS)$-th matrix block
is held by the processor labeled by 
\begin{equation}
	\Proc{\mathrm{mod}(\IS-1,\mathrm{Pr})\times\mathrm{Pc}+\mathrm{mod}(\JS-1,\mathrm{Pc})+1}.
	\label{}
\end{equation}
This is called a 2D block cyclic data-to-processor mapping.  The 
mapping itself does not take the sparsity of the matrix into account.
If the $(\IS,\JS)$-th block contains only zero elements, then that block is not
stored.  It is possible that some nonzero blocks may contain several
rows of zeros.  These rows are not stored either.
As an example, a 4-by-3 grid of processors is depicted in
Fig.~\subref*{fig:pmatrix_grid}.  
The mapping between the 2D supernode partition of a sparse matrix 
and the 2D
processor grid in Fig.~\subref*{fig:pmatrix_grid} is depicted in Fig.~\subref*{fig:pmatrix_layout}. 
Each supernodal block column of \Lmat is distributed among processors
that belong to a column of the processor grid.  Each processor may own
multiple matrix blocks.  For instance, the nonzero rows in the second 
supernode are owned by processors $\Proc{2}$ and $\Proc{5}$.
More precisely, $\Proc{2}$ owns two nonzero blocks, while $\Proc{5}$ is
responsible for one block. Note that these nonzero blocks are not necessarily contiguous
in the global matrix. %The distribution is similar for \Umat.
Though the nonzero structure of \Amat is not 
taken into account during the distribution, it has been shown in practice that 2D layouts
leads to higher scalability for both dense~\cite{Blackford1997} and sparse Cholesky 
factorization~\cite{RothbergGupta1994}.

In the current implementation, \pselinv contains an interface 
that is compatible with the \superlu software package.  
In order to allow \pselinv to be easily integrated with other
$LU$ factorization codes, we create some intermediate
sparse matrix objects to hold the distributed $L$ and $U$ factors.  Such
intermediate sparse matrix objects will be overwritten by matrix blocks
of $A^{-T}$ in the selected inversion process.  Each nonzero block
$L(\IS,\JS)$ is stored as follows. Diagonal blocks 
are always stored as dense matrices which includes both $L(\IS,\IS)$ and
$U(\IS,\IS)$. Nonzero entries of $L(\IS,\JS)$
($\IS>\JS$) are stored contiguously as a dense matrix in a column-major
order even though row indices associated with the stored matrix elements
are not required to be contiguous.  
Nonzero entries of within $U(\IS,\JS)$ ($\IS<\JS$) are also stored as a dense matrix in a
contiguous array in a column major order. The nonzero column indices
associated with the nonzeros entries in $U(\IS,\JS)$ are not required to
be continuous either.  We remark that for matrices with highly non-symmetric 
sparsity patterns, it is more efficient to store the upper triangular 
blocks using the skyline structure shown in~\cite{LiDemmel2003}.
However, 
we choose to use a simpler data layout because it allows level-3 BLAS (GEMM) 
to be used in the selected inversion process. 

\subsection{Computing selected elements of $A^{-T}$ within each
supernode in parallel} \label{subsec:block}

In this section, we detail how steps \ref{alg1.step3} to
\ref{alg1.step5} in
Alg.~\ref{alg:selinvlu} can be completed in parallel.
We perform step \ref{alg1.step2} of Alg.~\ref{alg:selinvlu} in a separate pass,
since the data communication required in this step is relatively simple.
The processor that owns the block $L_{\KS,\KS}$ broadcasts $L_{\KS,\KS}$ to
all other processors within the same column processor group owning
nonzero blocks in the supernode $\KS$. Each processor in that group
performs the triangular solve $\hat{L}_{\IS,\KS}= L_{\IS,\KS}
\left(L_{\KS,\KS}\right)^{-1}$ for each nonzero block contained in the set
$\CS$ defined in step \ref{alg1.step2} of the algorithm.  
Because $L_{\IS,\KS}$ is not used in the subsequent steps of selected
inversion once $\hat{L}_{\IS,\KS}$ has been computed, it is overwritten 
by $\hat{L}_{\IS,\KS}$.
Similarly, $U_{\KS,\KS}$ is broadcast to
all other processors within the same row processor group owning
nonzero blocks in the supernode $\KS$. Each processor in that group
performs the triangular solve $\hat{U}_{\KS,\IS} =
\left(U_{\KS,\KS}\right)^{-1}  U_{\KS,\IS} $ for each nonzero block contained in the set
$\CS$ defined in step \ref{alg1.step2} of the algorithm.

\begin{figure}[htbp]
\centering
\subfloat{
\begin{adjustbox}{scale=0.8}
\begin{tikzpicture}

    \pgfmathtruncatemacro{\Pr}{4};
    \pgfmathtruncatemacro{\Pc}{3};

\node at (13em, -17em) {\supernode{6}};
\node at (13em, -21.5em) {\supernode{8}};
\node at (13em, -27em) {\supernode{10}};

\node at (17em,-14em) {\supernode{6}};
\node at (21.5em,-14em) {\supernode{8}};
\node at (27em,-14em) {\supernode{10}};

%bounding box
%\draw[fill=none,draw=none] (9em,-13em) rectangle +(22em,-18em);

%D
\mapblock{6}{6}{\Pr}{\Pc}
\colorproc{\curproc}{\Pr}{\Pc}
\dblknew{\begin{topbox}{2em}{2em} \vspace{2pt} \center{\Proc{\curproc}} \end{topbox}}{2}{17}{,fill=proccolor};

\mapblock{8}{8}{\Pr}{\Pc}
\colorproc{\curproc}{\Pr}{\Pc}
\blk{\begin{topbox}{2em}{3em} \vspace{2pt} \center{\Proc{\curproc}} \end{topbox}}{2}{3}{22}{21.5}{,fill=proccolor};

\mapblock{10}{10}{\Pr}{\Pc}
\colorproc{\curproc}{\Pr}{\Pc}
\dblknew{}{4}{27}{,fill=proccolor};

%L
%6

%6 8 and 8 6
\mapblock{6}{8}{\Pr}{\Pc}
\colorproc{\curproc}{\Pr}{\Pc}
\blk{\begin{topbox}{2em}{3em} \vspace{2pt} \center{\Proc{\curproc}} \end{topbox}}{2}{3}{17}{21.5}{,fill=proccolor};
\mapblock{8}{6}{\Pr}{\Pc}
\colorproc{\curproc}{\Pr}{\Pc}
\blk{\begin{topbox}{2em}{2em} \vspace{2pt} \center{\Proc{\curproc}} \end{topbox}}{2}{2}{22}{17}{,fill=proccolor};

%\mlublk{}{6}{8}{2}{3}{17}{21.5}{\Pr}{\Pc};
\mapblock{6}{10}{\Pr}{\Pc}
\colorproc{\curproc}{\Pr}{\Pc}
\blk{\begin{topbox}{2em}{4em} \vspace{2pt} \center{\Proc{\curproc}} \end{topbox}}{2}{4}{17}{27}{,fill=proccolor};
\mapblock{10}{6}{\Pr}{\Pc}
\colorproc{\curproc}{\Pr}{\Pc}
\blk{\Proc{\curproc}}{4}{2}{27}{17}{,fill=proccolor};

%\elublkFill{}{2}{1}{19}{33.5};
%\elublkFill{}{2}{1}{19}{35.5};
%\node at (19em,-34em) {\Proc{13}};
%\node at (34em,-19em) {\Proc{6}};

%8
%\mlublk{+}{8}{10}{3}{4}{21.5}{27}{\Pr}{\Pc};

\mapblock{8}{10}{\Pr}{\Pc}
\colorproc{\curproc}{\Pr}{\Pc}
\blk{}{2}{4}{22}{27}{,fill=proccolor};
\mapblock{10}{8}{\Pr}{\Pc}
\colorproc{\curproc}{\Pr}{\Pc}
\blk{}{4}{3}{27}{21.5}{,fill=proccolor};

\mapblock{10}{10}{\Pr}{\Pc}
\colorproc{\curproc}{\Pr}{\Pc}
%\blk{}{4}{1}{27}{27.5}{,fill=proccolor!40,draw=none};
\blk{}{1}{4}{27.5}{27}{,fill=proccolor!40,draw=none};

\eublkFill{}{2}{1}{17}{27.5};

\eublkFill{}{3}{1}{21.5}{27.5};
\mapblock{10}{6}{\Pr}{\Pc}
\node at (27em,-17em) {\Proc{\curproc}};

\mapblock{8}{10}{\Pr}{\Pc}
\node at (21.5em,-25.7em) {\Proc{\curproc}};
\mapblock{10}{8}{\Pr}{\Pc}
\node at (27em,-20.6em) {\Proc{\curproc}};

\mapblock{10}{10}{\Pr}{\Pc}
\node at (27em,-25.7em) {\Proc{\curproc}};

%6
%bcasts of U
\draw[dashed,ultra thick, -latex] (22.5em, -16.5em) -| (23.8em, -16.5em) |- (22.5em,-21.5em);
\draw[dashed,ultra thick, -latex] (22.5em, -16.5em) -| (23.8em, -16.5em) |- (22.5em,-27em);

\draw[dashed,ultra thick, -latex] (28.5em, -16.5em) -| (30.8em, -16.5em) |- (28.5em,-21.5em);
\draw[dashed,ultra thick, -latex] (28.5em, -16.5em) -| (30.8em, -16.5em) |- (28.5em,-27em);

%bcasts of L 8,6
\draw[ultra thick, -latex] (17em , -22.5em ) |- (17em, -24em) -| (21.5em,-22.5em);
\draw[ultra thick, -latex] (17em , -22.5em ) |- (17em, -24em) -| (28.5em,-22.5em);

%bcasts of L 10,6
\draw[ultra thick, -latex] (17em , -28.5em ) |- (17em, -30.5em) -| (21.5em,-28.5em);
\draw[ultra thick, -latex] (17em , -28.5em ) |- (17em, -30.5em) -| (28.5em,-28.5em);

\node at (30.8 em, -18em) {\circled{b}};
\node at (23.8 em, -18em) {\circled{b}};

\node at (21.5 em, -22em) {\circled{1}};
\node at (26 em, -22em) {\circled{1}};
\node at (21.5 em, -27.5em) {\circled{1}};
\node at (26 em, -27.5em) {\circled{1}};

\node at (22.5 em, -22.5em) {\circled{2}};
\node at (27 em, -22.5em) {\circled{2}};
\node at (22.5 em, -28em) {\circled{2}};
\node at (27 em, -28em) {\circled{2}};

\node at (19.5 em, -30.5em) {\circled{a}};
\node at (19.5 em, -24em) {\circled{a}};

\end{tikzpicture}
\end{adjustbox}
}
~~~~\subfloat{
\begin{adjustbox}{scale=0.8}
\begin{tikzpicture}

    \pgfmathtruncatemacro{\Pr}{4};
    \pgfmathtruncatemacro{\Pc}{3};

\node at (12em, -17em) {\supernode{6}};
\node at (12em, -21.5em) {\supernode{8}};
\node at (12em, -27em) {\supernode{10}};

\node at (17em,-14em) {\supernode{6}};
\node at (21.5em,-14em) {\supernode{8}};
\node at (27em,-14em) {\supernode{10}};

%bounding box
%\draw[fill=none,draw=none] (9em,-13em) rectangle +(22em,-18em);

%D
\mapblock{6}{6}{\Pr}{\Pc}
\colorproc{\curproc}{\Pr}{\Pc}
\dblknew{\begin{topbox}{2em}{2em} \vspace{2pt} \center{\Proc{\curproc}} \end{topbox}}{2}{17}{,fill=proccolor};

\mapblock{8}{8}{\Pr}{\Pc}
\colorproc{\curproc}{\Pr}{\Pc}
\blk{\begin{topbox}{2em}{3em} \vspace{2pt} \center{\Proc{\curproc}} \end{topbox}}{2}{3}{22}{21.5}{,fill=proccolor};

\mapblock{10}{10}{\Pr}{\Pc}
\colorproc{\curproc}{\Pr}{\Pc}
\dblknew{}{4}{27}{,fill=proccolor};

%L
%6 8 and 8 6
\mapblock{6}{8}{\Pr}{\Pc}
\colorproc{\curproc}{\Pr}{\Pc}
\blk{\begin{topbox}{2em}{3em} \vspace{2pt} \center{\Proc{\curproc}} \end{topbox}}{2}{3}{17}{21.5}{,fill=proccolor};
\mapblock{8}{6}{\Pr}{\Pc}
\colorproc{\curproc}{\Pr}{\Pc}
\blk{\begin{topbox}{2em}{2em} \vspace{2pt} \center{\Proc{\curproc}} \end{topbox}}{2}{2}{22}{17}{,fill=proccolor};

\mapblock{6}{10}{\Pr}{\Pc}
\colorproc{\curproc}{\Pr}{\Pc}
\blk{\begin{topbox}{2em}{4em} \vspace{2pt} \center{\Proc{\curproc}} \end{topbox}}{2}{4}{17}{27}{,fill=proccolor};
\mapblock{10}{6}{\Pr}{\Pc}
\colorproc{\curproc}{\Pr}{\Pc}
\blk{\Proc{\curproc}}{4}{2}{27}{17}{,fill=proccolor};

%8
\mapblock{8}{10}{\Pr}{\Pc}
\colorproc{\curproc}{\Pr}{\Pc}
\blk{}{2}{4}{22}{27}{,fill=proccolor};
\mapblock{10}{8}{\Pr}{\Pc}
\colorproc{\curproc}{\Pr}{\Pc}
\blk{}{4}{3}{27}{21.5}{,fill=proccolor};

\mapblock{10}{10}{\Pr}{\Pc}
\colorproc{\curproc}{\Pr}{\Pc}
\blk{}{1}{4}{27.5}{27}{,fill=proccolor!40,draw=none};

\eublkFill{}{2}{1}{17}{27.5};
\eublkFill{}{3}{1}{21.5}{27.5};

\mapblock{10}{6}{\Pr}{\Pc}
\node at (27em,-17em) {\Proc{\curproc}};

\mapblock{8}{10}{\Pr}{\Pc}
\node at (21.5em,-25.7em) {\Proc{\curproc}};

\mapblock{10}{8}{\Pr}{\Pc}
\node at (27em,-20.6em) {\Proc{\curproc}};

\mapblock{10}{10}{\Pr}{\Pc}
\node at (27em,-25.7em) {\Proc{\curproc}};

%6
%reduce L
\draw[ultra thick, -latex] (28.5em,-20.6em) .. controls (30.5em,-20.6em) and (30.5em,-20.6em) .. (30.5em,-19.5em) .. controls (30.5em,-17.5em) and (30.5em,-17.5em) .. (28.5em, -17.5em);
\draw[ultra thick, -latex] (28.5em,-26em) .. controls (30.5em,-26em) and (30.5em,-26em) .. (30.5em,-19.5em) .. controls (30.5em,-17.5em) and (30.5em,-17.5em) .. (28.5em, -17.5em);

\draw[ultra thick, -latex] (22.5em,-20.6em) .. controls (24em,-20.6em) and (24em,-20.6em) .. (24em,-19.5em) .. controls (24em,-17.5em) and (24em,-17.5em) .. (22.5em, -17.5em);
\draw[ultra thick, -latex] (22.5em,-26em) .. controls (24em,-26em) and (24em,-26em) .. (24em,-19.5em) .. controls (24em,-17.5em) and (24em,-17.5em) .. (22.5em, -17.5em);

%reduce U
\draw[dashed,ultra thick, -latex] (21.5em,-22.5em) .. controls (21.5em,-24.5em) and (21.5em,-24.5em) .. (20em,-24.5em) .. controls (17em,-24.5em) and (17em,-24.5em) .. (17em, -22.5em);
\draw[dashed,ultra thick, -latex] (26em,-22.5em) .. controls (26em,-24.5em) and (26em,-24.5em) .. (20em,-24.5em) .. controls (17em,-24.5em) and (17em,-24.5em) .. (17em, -22.5em);

\draw[dashed,ultra thick, -latex] (21.5em,-28.5em) .. controls (21.5em,-30.5em) and (21.5em,-30.5em) .. (20em,-30.5em) .. controls (17em,-30.5em) and (17em,-30.5em) .. (17em, -28.5em);
\draw[dashed,ultra thick, -latex] (26em,-28.5em) .. controls (26em,-30.5em) and (26em,-30.5em) .. (20em,-30.5em) .. controls (17em,-30.5em) and (17em,-30.5em) .. (17em, -28.5em);

%reduce D from U
\draw[ultra thick, -latex] (16.5em,-21.6em) .. controls (14em,-21.6em) and (14em,-21.6em) .. (14em,-19.5em) .. controls (14em,-17.5em) and (14em,-17.5em) .. (16.5em, -17.5em);
\draw[ultra thick, -latex] (16.5em,-27em) .. controls (14em,-27em) and (14em,-27em) .. (14em,-19.5em) .. controls (14em,-17.5em) and (14em,-17.5em) .. (16.5em, -17.5em);

\node at (24.2 em, -19em) {\circled{c}};
\node at (30.8 em, -20em) {\circled{c}};

\node at (16.5 em, -22em) {\circled{3}};
\node at (16.5 em, -27.5em) {\circled{3}};

\node at (17 em, -18em) {\circled{4}};

\node at (19.5 em, -30.5em) {\circled{d}};
\node at (19.5 em, -24.5em) {\circled{d}};

\node at (14 em, -24.5em) {\circled{e}};

\end{tikzpicture}
\end{adjustbox}
}

\caption{
Communication and computational events for computing selected elements of $A^{-1}$ within \supernode{6}. 
The \circled{a}-\circled{1}-\circled{c} sequence of events yields $\{A_{i,6}^{-1} | i\in \CS_{U}\}$ and overwrites
the corresponding elements in $\hat{U}_{6,i}$.  The \circled{b}-\circled{2}-\circled{d} sequence yields 
$\{A_{6,j}^{-1} | j \in \CS_{L} \}$. Before overwriting the corresponding elements in $\hat{L}_{j,6}$, 
the \circled{3}-\circled{e}-\circled{4} sequence yields $A^{-1}_{6,6}$. 
Note that the shaded area of $A^{-T}_{10,10}$ does not contribute to supernode \supernode{6}.
}

\label{fig:step3}

\end{figure}
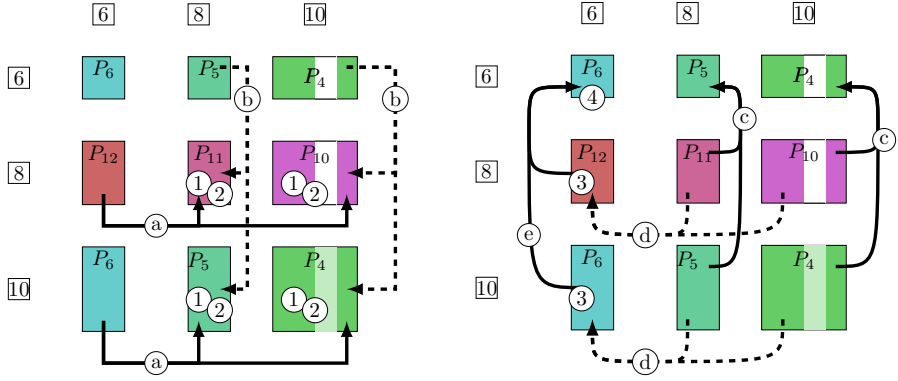

A more complicated communication pattern is required to complete steps
\ref{alg1.step3} to \ref{alg1.step5} in parallel.  Because
$A^{-T}_{\CL,\CU}$ and $\hat{L}_{\CL,\KS}$ (resp. $\hat{U}_{\KS,\CU}$)
are generally owned by different processor groups, using the approach
discussed in~\cite{JacquelinLinYang2017}, we need to send blocks of
$\hat{L}_{\CL,\KS}$ to processors that own {\em matching} blocks of
$A^{-T}_{\CL,\CU}$, so that matrix-matrix multiplication can be
performed on the group of processors owning $A^{-T}_{\CL,\CU}$.  More
specifically, the processor owning the $\hat{L}_{\IS,\KS}$ block sends
to all processors within the same row group of processors among which
$A^{-T}_{\IS,\CU}$ is distributed in step~\ref{alg1.step3}. 
%Similarly, in
%step~\ref{alg1.step5} the processor owning the $\hat{U}_{\KS,\JS}$ block
%sends to all processors within the same column group of processors among
%which $A^{-T}_{\CL,\JS}$ is distributed.

However, the set of processors owning $\hat{L}_{\IS,\KS}$ and
the owners of $A^{-T}_{\IS,\CU}$ generally form a small subset of all
processors, and this set can largely vary across different supernodes.
In order to perform such collective communication operations efficiently within the MPI framework, one would
have to create a communicator per distinct communication pattern.
We have shown in~\cite{JacquelinLinWichmannEtAl2016} that in the context of \pselinv, this
can result in more communicators than what was handled by most MPI
implementations for matrices of large sizes.
Therefore, one way to complete this step of data communication 
is to use a number of point-to-point asynchronous MPI sends from 
the processor that owns $\hat{L}_{\IS,\KS}$ to
the group of processors that own the nonzero blocks of $A^{-T}_{\IS,\CU}$. 
Similarly, in step~\ref{alg1.step5} the processor that owns $\hat{U}_{\KS,\JS}$ has to send it to 
the group of processors that own the nonzero blocks of $A^{-T}_{\CL,\JS}$.
Then $\hat{L}_{\IS,\KS}^{T} A^{-T}_{\IS,\JS}$ and $A^{-T}_{\IS,\JS} \hat{U}_{\KS,\JS}^{T}$
are performed locally on each processor owning $A^{-T}_{\IS,\JS}$ using the GEMM
subroutine in BLAS3,
and the local matrix contributions $\hat{L}_{\IS,\KS}^{T} A^{-T}_{\IS,\JS}$ are 
reduced within each column communication groups owning $\hat{U}_{\KS,\JS}$ to
produce the $A^{-T}_{\KS,\JS}$ block in step~\ref{alg1.step3} of Alg.~\ref{alg:selinvlu}. 
Respectively, local matrix contributions $A^{-T}_{\IS,\JS}\hat{U}_{\KS,\JS}^{T}$
are reduced within each row communication groups owning $\hat{L}_{\IS,\KS}$
to produce the $A^{-T}_{\IS,\KS}$ block in step~\ref{alg1.step5} of Alg.~\ref{alg:selinvlu}. 
We will discuss in more detail how these asynchronous point-to-point exchanges can be organized
to form efficient broadcast and reduction operations in
section~\ref{subsec:commun}.
%\LL{change the flow of the discussion}

Fig.~\ref{fig:step3} illustrates how this step is completed for a
specific supernode $\KS=\supernode{6}$, for the matrix
depicted in Fig.~\subref*{fig:pmatrix_layout}.  We use circled letters
$\circled{a}, \circled{b}, \circled{c}, \circled{d}, \circled{e}$ to label 
communication events, and circled numbers
$\circled{1}, \circled{2}, \circled{3}, \circled{4}$ to label computational events. 
We can see from this figure that $\hat{L}_{8,6}$ is sent by $P_{12}$ 
to all processors within the same row processor group to which $P_{12}$ 
belongs (\circled{a}). This group includes $P_{10}$, $P_{11}$, and $P_{12}$.  
Similarly $\hat{L}_{10,6}$ is broadcast from $P_6$ to all
other processors within the same row group to which $P_6$ belongs (\circled{a}).
For the upper triangular part, $\hat{U}_{6,8}$ is sent by $P_{5}$ along
the column processor group to which it belongs (\circled{b}). $P_4$ does a similar
communication operation for $\hat{U}_{6,10}$.

Local matrix-matrix multiplications are then performed on 
$P_{11}$, $P_{10}$, $P_{4}$ and $P_5$ simultaneously, corresponding 
to events \circled{1} and \circled{2}. Contributions to $A^{-T}_{\CL,\KS}$
are then reduced onto $P_{12}$ and $P_{6}$ within the row 
processor groups they belong to respectively (communication step \circled{d}).
Similarly, communication step \circled{c} corresponds to reductions of contributions
to $A^{-T}_{\KS,\CU}$ onto $P_5$ and $P_4$.
After this step, $A^{-T}_{8,6}$ and $A^{-T}_{10,6}$ become available on
$P_{12}$ and $P_{6}$ respectively.  The matrix product 
$\hat{L}_{\CL,\KS}^{T} A_{\CL,\KS}^{-T}$ 
is first
computed locally on the processor holding blocks of $\hat{L}_{\CL,\KS}$ (step \circled{3}),
and then reduced to the processor that owns the diagonal block $\hat{L}_{\KS,\KS}$
within the column processor group to which supernode $\KS$ is mapped (step \circled{e}).
The result of this reduction is added to the diagonal block during step \circled{4}. 
%step 5 of Alg.~\ref{alg:selinvlu}.
%As an example we use again Fig.~\ref{fig:step3} for $\KS=\supernode{6}$. 
%$\hat{U}_{6,8}^{T} A^{-T}_{6,8} \hat{L}_{8,6}^{T}$ is computed on $P_{12}$
%and sent to $P_{6}$.  Similarly $\hat{U}_{6,10}^{T} A^{-T}_{6,10} \hat{L}_{10,6}^{T}$
%is computed on $P_{6}$.
%Finally $P_{6}$ updates $A^{-T}_{6,6}$.
%Finally, $\hat{L}_{\CL,\KS}$ can be overwritten with $\hat{U}_{\KS,\CU}^{T} A_{\CL,\CU}^{-T}$
%and $\hat{U}_{\KS,\CU}$ with $A_{\CL,\CU}^{-T}\hat{L}_{\CL,\KS}^{T}$.
This completes the computation for the current supernode $\KS$, and the
algorithms moves to the next supernode. 
%$(\KS-1)$.
%
%After step $6$, the algorithm .

\subsection{Task scheduling and asynchronous collective
communication\label{subsec:commun}}

%In this section, we describe the strategy we have used in the implementation of
%the non-symmetric \pselinv algorithm to achieve high strong scalability.
%
%As it is widely known, it is extremely important for nowadays implementation to overlap
%communications with computations. Indeed, the relative cost of communications is getting
%higher as the gap between processing speed and network bandwidth/latency is widening.
%
%We have spent a lot of effort on this topic in the context of the symmetric 
%Selected Inversion implementation presented in~\cite{REFERENCE}. In addition to 
%exploiting parallelism within a supernode, as described in section~\ref{subsec:block} 
%for non-symmetric matrices, there is potentially a large amount of concurrency between
%different supernodes which can be exploited too. We have proposed to organize the
%computations by scheduling supernodes sharing the same level in the elimination tree
%at the same step, allowing for concurrent processing. In our implementation of \pselinv
%for non-symmetric matrices, we use the same strategy and encourage the interested reader
%to refer to our previous report~\cite{REFERENCE} for additional details.

In section~\ref{subsec:block}, we have discussed how to exploit parallelism
within a given supernode. Besides such intra-node parallelism, there is
potentially a large amount of inter-node concurrency across the work
associated with different supernodes. In~\cite{JacquelinLinYang2017} we
have demonstrated that exploiting such inter-node parallelism is crucial
for improving the parallel scalability of the \pselinv method for
symmetric matrices. The basic idea is to use the elimination
tree~\cite{Liu1990}
associated with the sparse $LU$ factorization to add an additional coarse-grained level 
of parallelism at the {\tt for} loop level in Alg.~\ref{alg:selinvlu}.
For non-symmetric matrices we use the same strategy to exploit the
inter-node parallelism.

We create a basic parallel task scheduler to launch different iterates of the 
{\tt for} loop in a certain order. This order is defined by
a priority list $S$, which is indexed by integer priority numbers 
ranging from 1 to $n_s$, where
$n_s$ is bounded from above by the depth of the elimination tree.
The task performed in each iteration of the {\tt for} loop is assigned 
a priority number $\sigma(\mathcal{I})$.  The lower the number, the higher the 
priority of the task, hence the sooner it is scheduled. 
The supernode $\mathcal{N}$ associated with the root of the elimination tree 
clearly has to be processed first. 
If multiple supernodes or tasks have the same priority number, they are executed in a random order.
Even though we use a priority list to help launch tasks, 
we do not place extra synchronization among launched tasks other than 
requiring them to preserve data dependency. 
Tasks associated with different supernodes can be executed 
concurrently if these supernodes are on different critical paths 
of the elimination tree, and if there is no overlap among processors mapped to 
these critical paths.
%In fact, if tasks associated with supernode $\mathcal{J}$ and $\mathcal{I}$ are mapped to 
%different sets of processors, the task associated with the supernode $\mathcal{J}$ may
%actually start before that associated with another supernode $\mathcal{I}$
%even if $\sigma(\mathcal{I}) < \sigma(\mathcal{J})$,
% i.e. even if task $\mathcal{I}$ is
%scheduled ahead of task $\mathcal{J}$ according to the priority list.
%When two different tasks need to communicate
%with a common ancestor, the priority number associated with each task
%determines which task is completed first. 
We refer readers
to~\cite{JacquelinLinYang2017} for more details on how to
create such a task scheduler. 

\begin{figure}[htbp]
\centering
\begin{adjustbox}{width=.5\linewidth}
\begin{tikzpicture}

\draw[draw=none] (-7.8,1.5) rectangle +(10.6,-13);

\begin{scope}[yscale=-1]
\draw[draw=none] (-2.8,-1.5) rectangle +(7.5,13);

\node[fill=red!40,   ultra thick,circle,minimum width = 3em,draw=black] (pr) at (0,0) {$P_4$};
\node[fill=blue!40,  ultra thick,circle,minimum width = 3em,draw=black] (po) at (0,2) {$P_1$};
\node[fill=green!40, ultra thick,circle,minimum width = 3em,draw=black] (pq) at (0,4) {$P_2$};
\node[fill=orange!40,  ultra thick,circle,minimum width = 3em,draw=black] (ps) at (0,6) {$P_3$};
\node[fill=purple!40,ultra thick,circle,minimum width = 3em,draw=black] (pt) at (0,8) {$P_5$};
\node[fill=brown!40, ultra thick,circle,minimum width = 3em,draw=black] (pn) at (0,10) {$P_6$};

\path (pr.west) edge[thick,out=-220,in=-160,looseness =0.6,-latex]  (pn.west);
\path (pn.east) edge[thick,out=40,in=-20,looseness =0.8,-latex]  (po.east);
\path (pn.east) edge[thick,out=-40,in=20,looseness =0.65,-latex]  (pq.east);
\path (pr.west) edge[thick,out=-220,in=-160,looseness =0.55,-latex]  (ps.west);
\path (ps.east) edge[thick,out=20,in=-20,looseness =0.8,-latex]  (pt.east);
\end{scope}

\begin{scope}[yscale=-1, xshift=-2em]

\draw[ultra thick] (-9em,4.8) -- (-6em,4.8) -- (-6em,4.6) -- (-5em,5) -- (-6em,5.4) -- (-6em,5.2) -- (-9em,5.2) -- (-9em,4.8);

\end{scope}

\begin{scope}[yscale=-1, xshift=-19em]
\draw[draw=none] (-2.8,-1.5) rectangle +(7.5,13);
%\draw[ultra thick,draw=none] (7.5,0) -- (7.5,10);

\node[fill=red!40,   ultra thick,circle,minimum width = 3em,draw=black] (pr) at (0,0) {$P_4$};
\node[fill=blue!40,  ultra thick,circle,minimum width = 3em,draw=black] (po) at (0,4) {$P_1$};
\node[fill=green!40, ultra thick,circle,minimum width = 3em,draw=black] (pq) at (0,6) {$P_2$};
\node[fill=orange!40,  ultra thick,circle,minimum width = 3em,draw=black] (ps) at (0,8) {$P_3$};
\node[fill=purple!40,ultra thick,circle,minimum width = 3em,draw=black] (pt) at (0,10) {$P_5$};
\node[fill=brown!40, ultra thick,circle,minimum width = 3em,draw=black] (pn) at (0,2) {$P_6$};

\draw (-2,1) -- (2,1);
\draw[dashed] (-2,7) -- (2,7);
\draw (-1,3) -- (1,3);
\draw[dashed] (-1,5) -- (1,5);

\draw (-1,9) -- (1,9);

\path (pr.east) edge[thick,out=40,in=-20,looseness =0.9,-latex]  (pn.east);
\path (pn.east) edge[thick,out=40,in=-20,looseness =0.8,-latex]  (po.east);
\path (pn.east) edge[thick,out=40,in=-20,looseness =0.65,-latex]  (pq.east);
\path (pr.east) edge[thick,out=40,in=-20,looseness =0.55,-latex]  (ps.east);
\path (ps.east) edge[thick,out=20,in=-20,looseness =0.8,-latex]  (pt.east);

\end{scope}

\end{tikzpicture}
\end{adjustbox}
\caption{A random shifted binary tree broadcast: ranks are randomly shifted before organizing the broadcast along a binary tree.\label{fig.modbtree}}
\end{figure}
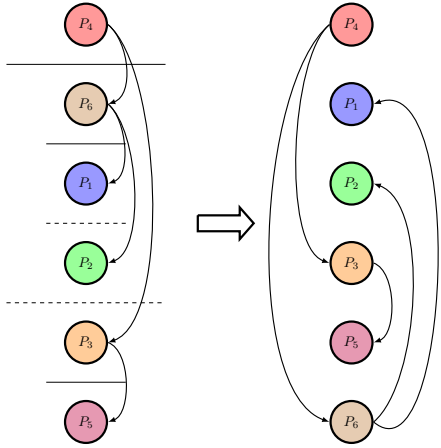

Collective communication operations such as broadcast and reduction in
section~\ref{subsec:block} dominate the communication cost of the
\pselinv method. Each communication events involves potentially a
different group of processors, and it is not practical to create an MPI
communicator per group especially when a large number of processors are
used.
Instead, our implementation relies on asynchronous point-to-point
MPI\_Isend/MPI\_Irecv routines to communicate between the processors.
Take the broadcast operation for example, 
the simplest strategy is to let one processor to send information to all
other processors within the relevant communication group. However, such
a simple strategy can result in a highly imbalanced communication
volume, as demonstrated in~\cite{JacquelinLinWichmannEtAl2016} for symmetric
matrices. Instead, we employ the \textit{shifted binary tree} method developed
in~\cite{JacquelinLinWichmannEtAl2016} for asynchronous communication
operations. Assuming that ranks are sorted, this type of tree is built 
by first shifting ranks of the recipients around a random position, and then by
building a binary tree from the root to those shifted ranks.
An example of a such tree depicted in Fig.~\ref{fig.modbtree}.

In the non-symmetric implementation of \pselinv, we therefore use non-blocking random shifted 
binary trees for the 
following operations:
\begin{enumerate}
\item broadcasting $\hat{L}_{\CL,\KS}$ to processors owning $A^{-T}_{\CL,\CU}$ (step \circled{a}),
\item broadcasting $\hat{U}_{\KS,\CU}$ to processors owning $A^{-T}_{\CL,\CU}$ (step \circled{b}),
\item reducing contributions to $A^{-T}_{\KS,\CU}$ (step \circled{c}),
\item reducing contributions to $A^{-T}_{\CL,\KS}$ (step \circled{d}),
\item reducing contributions to $A^{-T}_{\KS,\KS}$ (step \circled{e}).
\end{enumerate}

%The same approach has been used with success in the implementation of \pselinv for 
%symmetric matrices, as described in~\cite{JacquelinLinWichmannEtAl2016}.

\section{Numerical results}\label{sec:numerical}

%\LL{Examples to show: 
%\begin{itemize}
%  \item For instance, 1 symmetric matrix market matrix, 1 n
%    structurally symmetric matrix market matrix, 2 fully
%    asymmetric matrix market matrices. One SIESTA DNA matrices from k-point
%    sampling, an 3 SIESTA C-BN matrices (for weak scaling tests)
%  \item Compare the symmetric and asymmetric version of \pselinv for the
%    symmetric matrix in terms of a) wall clock time b) memory
%    consumption.  Both uses the tree level parallelization directly, and
%    maybe add a line for asymmetric version without tree level
%    parallelization.
%  \item Strong scaling,  and use the SIESTA matrices to show weak scaling .
%  \item Accuracy, take two matrices and compare with MUMPS. 
%  \item Take two matrices and compare the efficiency with MUMPS.
%\end{itemize}
%}

We evaluate the performance of \pselinv on a
variety of problems, taken from sources including the University
of Florida Matrix Collection\cite{FloridaMatrix}, and matrices generated
from the SIESTA~\cite{SolerArtachoGaleEtAl2002} and
DGDFT~\cite{LinLuYingE2012}, two software packages for performing Kohn-Sham
density functional theory~\cite{KohnSham1965} calculations using two different types of
basis sets.  
The first matrix collection is a widely used benchmark set of problems for
testing sparse direct methods, while the other set comes from
practical large scale electronic structure calculations.  The names
of these matrices as well as some of their characteristics are listed in
Tables~\ref{tab:testproblem} and \ref{tab:characteristics}.  The
matrices labeled by SIESTA\_XXX\_k are obtained from the SIESTA package
with k-point sampling. These matrices are complex structurally symmetric
matrices, but are neither complex symmetric nor Hermitian.   The matrices labeled by DG\_XXX
and by SIESTA\_XXX are complex symmetric matrices. We include these
matrices in the test that compare
the performance of the non-symmetric \pselinv solver with that of \pselinv for symmetric matrices.

%The first
%three problems in these tables come with two matrices each. One of the 
%matrices, denoted by $H$, is a discretized Hamiltonian, and the other matrix
%is an overlap matrix denoted by $S$. For all other problems, the overlap
%matrices can be considered as the identity matrix.
%All matrices are real and symmetric. In all our experiments, we compute
%the selected elements of the matrix
%\begin{equation}
%A(z) = H - z S.
%\label{eq:az}
%\end{equation}
%%For simplicity, we choose $z=0$ for all the efficiency tests in
%%section~\ref{subsec:scalability}.  
%The $LU$ factorization is performed
%by using the \superlu software package.  
%\superlu does not use dynamic pivoting strategies and our matrices are permuted
%without taking into account the values of matrix entries. Consequently
%the efficiency of both \superlu and \pselinv is independent of the
%choice of $z$.  The lack of dynamic pivoting strategies may impact the
%accuracy of \pselinv for highly indefinite and near-singular systems.
%We study the accuracy for different choices of complex shifts $z$ in
%section~\ref{subsec:accuracy}.  All the timing results reported
%are performed in complex arithmetic computation.
%

\begin{table}[htbp]
  \centering
  \begin{tabular}{l|l}
    \toprule
    Problem & Description \\
    \midrule
    SIESTA\_Si\_512\_k & KSDFT, Si with 512
    atoms (complex structurally symmetric)\\
    SIESTA\_DNA\_25\_k & KSDFT, DNA with 17875
    atoms (complex structurally symmetric)\\
    SIESTA\_DNA\_64\_k & KSDFT, DNA with 45760
    atoms (complex structurally symmetric)\\
    SIESTA\_CBN\_0.00\_k & KSDFT, C-BN sheet with 12770 atoms
    (structurally symmetric)\\
    SIESTA\_Water\_4x4x4\_k & KSDFT, Water with 12288 
    atoms (complex structurally symmetric)\\
    \hline
    audikw\_1& \begin{tabular}{@{}l}Automotive crankshaft model with over 900,000 TETRA elements \\(real symmetric)\end{tabular}\\
    shyy161& \begin{tabular}{@{}l}Direct, fully-coupled method for solving 
the Navier-Stokes equations\\for viscous flow calculations (real non-symmetric)\end{tabular}\\
    stomach& Electro-physiological model of a Duodenum (real non-symmetric)\\
    \hline
    DG\_DNA\_715\_64cell& KSDFT,  DNA with 45760
    atoms (complex symmetric)\\
    DG\_Graphene8192& KSDFT, Graphene sheet with 8192 atoms (complex symmetric)\\
    \hline
		SIESTA\_C\_BN\_1x1 & KSDFT, C-BN sheet with
		2532 atoms (complex symmetric)\\ 
		SIESTA\_C\_BN\_2x2 & KSDFT, C-BN sheet with
		10128 atoms (complex symmetric)\\ 
		SIESTA\_C\_BN\_4x2 & KSDFT, C-BN sheet with
		20256 atoms (complex symmetric)\\ 
    \bottomrule
  \end{tabular}
  \caption{Description of test problems for \pselinv.}
  \label{tab:testproblem}
\end{table}

\begin{table}[htbp]
  \centering
  \begin{tabular}{l|c|c|c}
    \toprule
    problem & $n$ & $|A|$ & $|L+U|$ \\
    \midrule
    	SIESTA\_Si\_512\_k &  6,656 &    5,016,064 &   32,686,104 \\
      SIESTA\_DNA\_25\_k &179,575 &   87,521,775 &  351,534,751\\
      SIESTA\_DNA\_64\_k &459,712&  224,055,744&  904,281,098\\
      SIESTA\_CBN\_0.00\_k &166,010 &  251,669,372 & 2,907,670,098\\
      SIESTA\_Water\_4x4x4\_k & 94,208 &   32,706,432 & 1,388,275,840\\
    \hline
    audikw\_1&943,695&   77,651,847& 2,530,341,547\\
    shyy161&76,480&     329,762&    4,467,806\\
    stomach&213,360&    3,021,648&   83,840,514\\
    \hline
    DG\_DNA\_715\_64cell & 459,712&  224,055,744&  898,749,546\\
    DG\_Graphene8192 &      327,680&  238,668,800& 1,968,211,450\\
    \hline
		SIESTA\_C\_BN\_1x1 & 32,916 & 23,857,418 &  274,338,850\\
		SIESTA\_C\_BN\_2x2 & 131,664 & 95,429,672 &  1,655,233,542\\
		SIESTA\_C\_BN\_4x2 & 263,328  & 190,859,344  & 3,591,750,262 \\
%    DNA\_715\_64cell & 459,712& 224,055,744& 866,511,698\\
%		DG\_Graphene\_2048 & 82,944  & 87,340,032  &545,245,344 \\
%		DG\_Graphene\_8192 & 331,776  & 349,360,128  & 2,973,952,468\\
%    pwtk & 217,918 & 5,926,171 & 104,644,472  \\
%    parabolic\_fem & 525,825 & 3,674,625 & 58,028,731\\
%    ecology2 & 999,999 & 2,997,995 & 91,073,583 \\
%    audikw\_1 & 943,695& 77,651,847& 2,500,489,909\\
    \bottomrule
  \end{tabular}
  \caption{The dimension $n$, the number of nonzeros $|A|$, and the
  number of nonzeros of the factors $|L+U|$ of the test problems.}
  \label{tab:characteristics}
\end{table}

%
%To assess the performance of \pselinv, we conducted a 
%number of computational experiments which we report in this section.
%
%Our asymmetric test problems are taken from various sources including Harwell-Boeing Test Collection
%\cite{HarwellBoeing}, the University of Florida Matrix
%Collection\cite{FloridaMatrix}, and matrices generated from electronic
%structure software including SIESTA~\cite{SolerArtachoGaleEtAl2002} and
%DGDFT~\cite{LinLuYingE2012}. 
%The first two matrix collections are widely used benchmark problems for
%testing sparse direct methods, while the other test problems come from
%practical large scale electronic structure calculations.  The names
%of these matrices as well as some of their characteristics are listed in
%Tables~\ref{tab:testproblem} and \ref{tab:characteristics}. 

In all of our experiments, we used the NERSC Edison platform with Cray
XC30 nodes. Each node has 24 cores partitioned among two
Intel Ivy Bridge processors.  Each 12-core processor runs at 2.4GHz. A
single node has 64GB of memory, providing more than 2.6 GB of memory per
core. We run one MPI rank per core as an efficient multithreaded scheme
is not yet available in \pselinv implementation. 
Computations are performed in complex arithmetic for all packages.
Sparse matrices were reordered to reduce the amount of fill using PARMetis 4.0.3~\cite{KarypisKumar1998}
in all experiments. 
Before applying \pselinv, a $LU$ factorization is first computed using \superlu 5.1.0.
In section~\ref{sec:mumps_comp}, we compare \pselinv to the \mumps
5.0.0~\cite{mumps,AmestoyDuffLExcellentEtAl2012,AmestoyDuffLExcellentEtAl2015} package to demonstrate the accuracy as well as the
efficiency of our implementation.

\subsection{Strong scaling experiments}\label{subsec:strongscale}

We illustrate the strong scalability of  \pselinv using several
non-symmetric and symmetric matrices.  In the latter case, the
non-symmetric storage format is used and performance is
compared against the symmetric implementation of \pselinv presented
in~\cite{JacquelinLinYang2017,JacquelinLinWichmannEtAl2016} and available in the \pexsi package\footnote{version 0.10.1 on \url{http://www.pexsi.org/}}.  Each experiment is repeated 10 times and the average timing measurements are
reported, together with error bars representing standard deviations in
the plots. 

Factorization timing measurements from \superlu are provided as a reference. $LU$ 
factorization is a pre-processing step of \pselinv, and needs
to be added to the selected inversion time to reflect the overall cost 
required to compute the selected elements of the inverse matrix.
Moreover, $LU$ factorization and selected inversion have the same
asymptotic computational cost but the actual cost may differ in practice.
For the SIESTA\_C\_BN\_2x2 matrix for instance, the $LU$
factorization requires $1.78373 \times 10^{13}$ floating point
operations (flops). The selected
inversion requires $3.59698 \times 10^{13}$ flops, which is around
$2$ times larger. This needs to be taken into 
consideration when comparing the factorization times to the selected inversion 
times.

\begin{figure}[htbp]
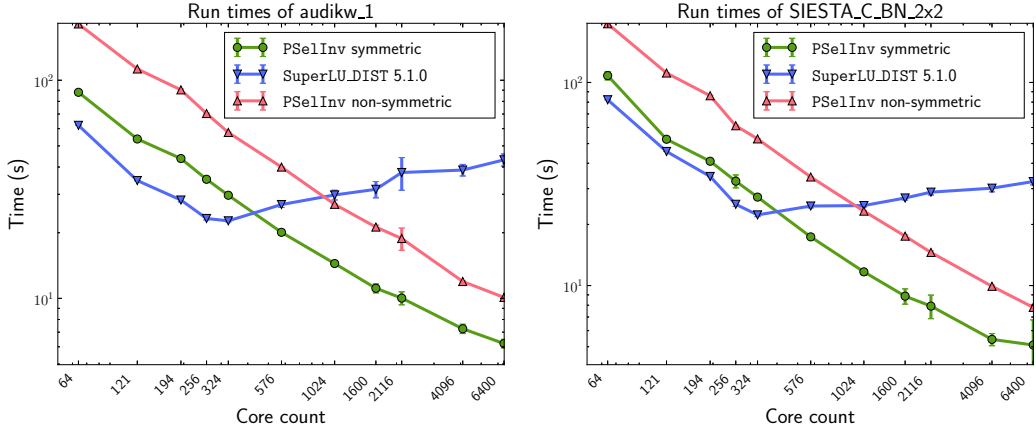

\centering
\subfloat{
\begin{adjustbox}{width=.45\linewidth}
\input{ss_audikw_1.pgf}
\end{adjustbox}
}
\subfloat{
\begin{adjustbox}{width=.45\linewidth}
\input{ss_LU_C_BN_C_2by2.pgf}
\end{adjustbox}
}
\caption{Strong scaling of \pselinv on audikw\_1 and SIESTA\_C\_BN\_2x2
matrices\label{fig.ss_audi_siesta2x2} }
\end{figure}

\begin{figure}[htbp]
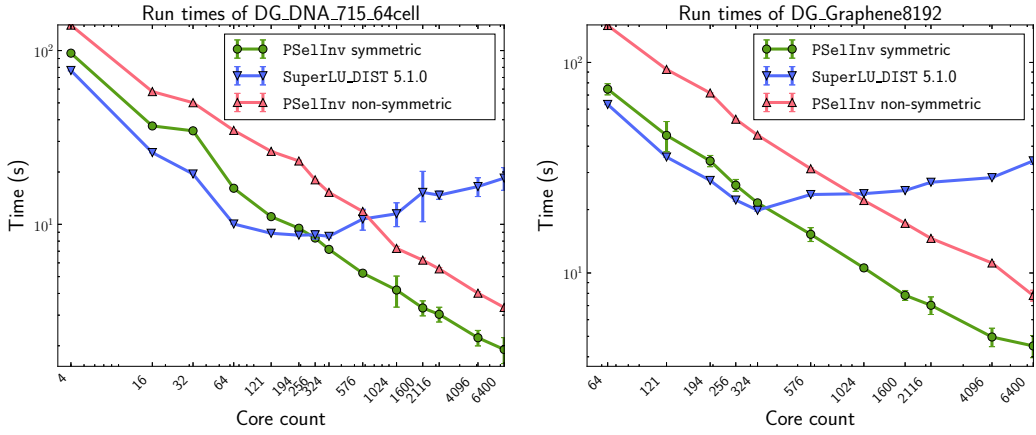

\centering
\subfloat{
\begin{adjustbox}{width=.45\linewidth}
\input{ss_DNA_715.pgf}
\end{adjustbox}
}
\subfloat{
\begin{adjustbox}{width=.45\linewidth}
\input{ss_DG_Graphene8192.pgf}
\end{adjustbox}
}
\caption{Strong scaling of \pselinv on DG\_DNA\_64 and DG\_Graphene8192 matrices\label{fig.ss_dg_dna_64_dg_graph_8192}}
\end{figure}

The first set of experiments (Fig.~\ref{fig.ss_audi_siesta2x2} and
Fig.~\ref{fig.ss_dg_dna_64_dg_graph_8192}) demonstrate that the strong
scalability of the non-symmetric version of \pselinv rivals that of the
symmetric version. Over these 4 matrices, \pselinv can scale up to 6,400
cores.  We also note that  \superlu can scale up to only 256 processors.
Based on the study in~\cite{JacquelinLinYang2017}, the scalability of
\pexsi greatly benefits from the strategy for handling collective
communication operations as well as the coarse-grain level parallelism.
The runtime of the non-symmetric version of \pselinv is 1.5--2.1 times
of that of the symmetric version, which illustrates the efficiency of
the non-symmetric implementation despite the more complex communication
pattern. In particular, we observe that such ratio tends to be smaller
than $2.0$ when more than $2000$ cores are used. This is because we have
removed some redundant data communication in the non-symmetric
implementation of \pselinv, and we plan to pursue such improved
implementation for the symmetric version of \pselinv in the future as
well.

\begin{figure}[htbp]
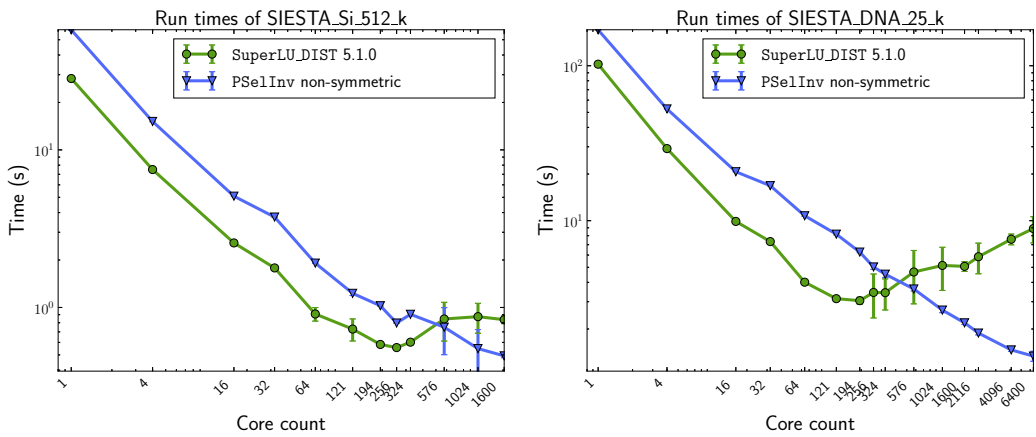

\centering
\subfloat{
\begin{adjustbox}{width=.45\linewidth}
\input{ss_Si_512.pgf}
\end{adjustbox}
}
\subfloat{
\begin{adjustbox}{width=.45\linewidth}
\input{ss_DNA_25.pgf}
\end{adjustbox}
}
\caption{Strong scaling of \pselinv on SIESTA\_Si\_512\_k and SIESTA\_DNA\_25\_k matrices\label{fig.ss_si_512_dna_25}}
\end{figure}

\begin{figure}[htbp]
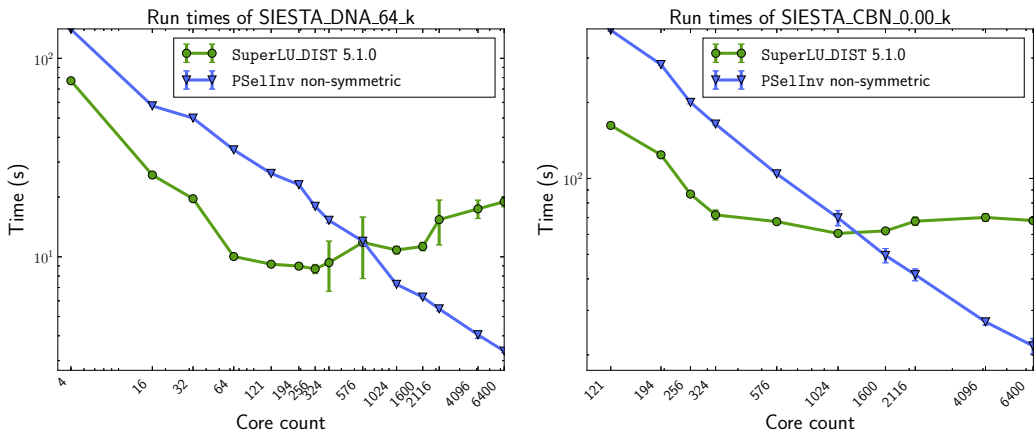

\centering
\subfloat{
\begin{adjustbox}{width=.45\linewidth}
\input{ss_DNA_64.pgf}
\end{adjustbox}
}
\subfloat{
\begin{adjustbox}{width=.45\linewidth}
\input{ss_CBN_0.00.pgf}
\end{adjustbox}
}
\caption{Strong scaling of \pselinv on SIESTA\_DNA\_64\_k and SIESTA\_CBN\_0.00\_k matrices\label{fig.ss_dna_64_cbn_0.00}}
\end{figure}

\begin{figure}[htbp]
\centering
\begin{adjustbox}{width=.45\linewidth}
\input{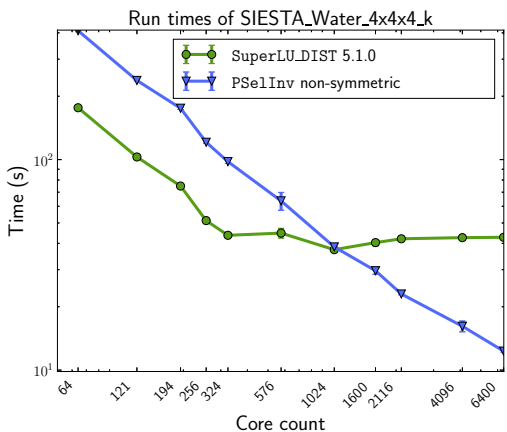}
\end{adjustbox}
\caption{Strong scaling of \pselinv on SIESTA\_Water\_4x4x4\_k matrix\label{fig.ss_water_4x4x4}}
\end{figure}

The next set of experiments focuses on assessing the 
efficiency of the \pselinv for the SIESTA\_XXX\_k matrices, which are
only structurally symmetric. These matrices corresponds to electronic
structure calculations of 1D, 2D and 3D quantum systems. This results in
the large difference in the ratio $|L+U|/|A|$ for different matrices. We
also stress that we do not explicitly take advantage of the structural
symmetry of the matrix.  The results depicted in
Fig.~\ref{fig.ss_si_512_dna_25}, Fig.~\ref{fig.ss_dna_64_cbn_0.00} and
Fig.~\ref{fig.ss_water_4x4x4} demonstrates that the performance of
\pselinv for non-symmetric matrices is comparable to that for symmetric
matrices. \pselinv can scale to up to 6,400 cores on all problems except the SIESTA\_Si\_512\_k matrix, which is
significantly smaller in size. On the other hand, \superlu can only scale
to around $300$ processors.

%These experiments show that non-symmetric \pselinv has good strong scalability on a wide collection of problems. 
%Its scalability is even higher than that of reference 
%packages such as \superlu, demonstrating the efficiency of the
%scheduling and communication strategies employed in \pselinv.

\subsection{Weak scaling experiment on symmetric matrices}

In this section we evaluate the weak scalability of the non-symmetric
version of \pselinv. Since the workload, measured by the flops  of \pselinv, generally does not scale linearly with
respect to the matrix size, we perform weak scaling tests by keeping the
flops per core close to be constant while increasing the matrix size
and the number of processors simultaneously.  We choose the
SIESTA\_C\_BN\_XXX matrices for demonstrating both the weak scaling and the computational complexity of \pselinv. These matrices 
correspond to electronic structure calculations of two dimensional C-BN
sheets of increasing sizes. 
%From the perspective of asymptotic complexity,
%the structure of these matrices can be viewed to resemble that of the
%two dimensional Laplace operator discretized by a finite difference
%scheme. 
For such matrices, asymptotic complexity
analysis~\cite{LinLuYingCarE2009} shows that the flop count should increase
by a factor of 8 from SIESTA\_C\_BN\_1x1 to SIESTA\_C\_BN\_2x2, but only by a factor of 2 from SIESTA\_C\_BN\_2x2 to
SIESTA\_C\_BN\_4x2, respectively.  The nonlinear growth behavior can be
explained in terms of the size of the largest separator
of the graph associated with the sparsity pattern of the matrix.
In the former case, the size of the largest separator
increases by a factor of $2$. The dense matrix inversion corresponding
to this separator leads to a factor of $2^3=8$ increase in flops.
In the latter case, the size of the
largest separator remains approximately the same despite the
increase of the matrix size. Hence the flops approximately increases
linearly with respect to the matrix size.
Table~\ref{tab.ws_conf} shows that the actual flop count
obtained from \pselinv agrees well with the theoretical
prediction: From SIESTA\_C\_BN\_1x1 to SIESTA\_C\_BN\_2x2 the flops
increase by a factor of 8.1, while an increase by a factor of 2.3 is seen from SIESTA\_C\_BN\_2x2 to
SIESTA\_C\_BN\_4x2. We choose the number of
cores so that the number of flops per core is approximately $5\times
10^{9}$.  The largest number of cores we used for this test is $576$ processors. This is due to the 
limitation of the strong scalability of \superlu as observed in
section~\ref{subsec:strongscale}.

%\todo[inline]{spy plot of two of them ? }

% We aim at keeping the workload per core constant. In order to do so, we first count the number of floating point operations required to perform the selected inversion on all three matrices. We then chose a value
% of approximately  , leading to the configurations 
% presented in Table~\ref{tab.ws_conf}.

\begin{table}[htbp]
\centering
\begin{tabular}{|l|c|c|c|}
\hline
Problem & $P$ & $flops$ & $flops / P$ \\
\hline
SIESTA\_C\_BN\_1x1 & 30 & $1.6\times 10^{13}$ &  $533\times 10^{9}$\\
SIESTA\_C\_BN\_2x2 & 256 & $1.3\times 10^{14}$ & $508\times 10^{9}$\\
SIESTA\_C\_BN\_4x2 & 576 & $3.0\times 10^{14}$ & $520\times 10^{9}$\\
\hline
\end{tabular}
\caption{Configurations used in the weak scaling experiments\label{tab.ws_conf}}

\end{table}

\begin{figure}[htbp]
\centering
\begin{adjustbox}{width=.7\linewidth}
\input{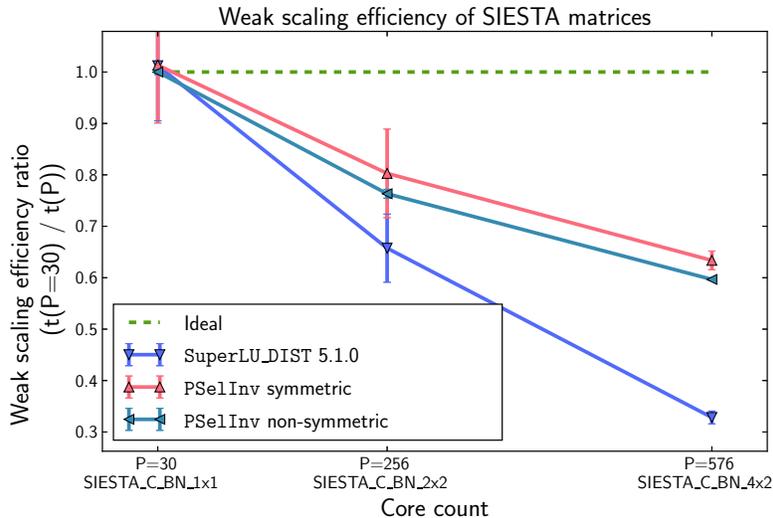}
\end{adjustbox}
\caption{Weak scaling of \pselinv and \superlu on SIESTA sparse matrices\label{fig.ws_siesta}}
\end{figure}

Fig.~\ref{fig.ws_siesta} shows that the non-symmetric
implementation of \pselinv exhibits similar weak scalability compared to that
of the symmetric case.  We again
repeat each experiment 10 times and report the averaged timing results, while
error bars represent standard deviations.  The line labeled by the
``ideal'' weak scaling is constructed by using the timing measurements obtained from a 30-core run.  We observe that that both the symmetric and
the non-symmetric versions of \pselinv exhibit better weak scalability
than that of  $LU$ factorization implemented in \superlu. The
non-symmetric version achieves weak scaling efficiency of 59\% on 576
cores, while the weak scaling efficiency of the symmetric version of \pselinv is slightly higher at
63\%. The weak scaling efficiency of \superlu is 33\% when $576$ cores
are used. 
%This demonstrate the efficiency of the
%communication strategy implemented in \pselinv.

\subsection{Comparison against the \mumps state-of-the-art
solver}\label{sec:mumps_comp}

In this section, we provide a comparative study of the performance of
the non-symmetric implementation of \pselinv against that of \mumps (version 5.0.0), which is a state-of-the-art sparse matrix solver. 
In addition to $LU$ factorization, the \mumps package also offers an 
optimized algorithm for solving multiple sparse right-hand sides which can be 
used to perform selected inversion as
well~\cite{AmestoyDuffLExcellentEtAl2012,AmestoyDuffLExcellentEtAl2015}.
This approach is more generic than the one presented in this paper 
which is more restrictive on the element selection in the matrix inverse. Similarly to \pselinv, \mumps first need to compute
the $LU$ factorization prior to computing the entries of the inverse.
In the following, we use \mumps to compute only the diagonal elements of
the inverse matrix, while \pselinv computes all entries corresponding to
Eq.~\ref{eq.sel_entries} including the diagonal elements.  Each
experiment is repeated 5 times and average times are reported.

\begin{figure}[htbp]
\centering
\subfloat{
\begin{adjustbox}{width=.45\linewidth}
\input{ss_MUMPS_shyy161.pgf}
\end{adjustbox}
}
\subfloat{
\begin{adjustbox}{width=.45\linewidth}
\input{ss_MUMPS_stomach.pgf}
\end{adjustbox}
}
\caption{Strong scaling of \pselinv and \mumps 5.0.0 on shyy161 and stomach matrices\label{fig.ss_stomach_shyy161}}
\end{figure}

The results in Fig.~\ref{fig.ss_stomach_shyy161} demonstrate that
\pselinv can be  orders of magnitude faster than the inversion available in
\mumps, even though \mumps computes only diagonal elements of the inverse.
The speedup achieved by \pselinv over \mumps inversion
reaches 27 for the shyy161 matrix, and 67 for the stomach matrix. 
Table~\ref{tab.accuracy} illustrates the accuracy of \pselinv is
fully comparable to that of \mumps, measured in terms of  
the diagonal entries of the matrix inverse.
We remark that the shyy161 matrix, the diagonal contains 
elements with very small magnitude (some are zero elements). Therefore,
row pivoting has to be used to move these elements to off-diagonal
positions. \superlu uses a static row pivoting strategy, while 
\mumps employs a dynamic one.  Table~\ref{tab.accuracy} shows that for
the matrix we tested, the static row pivoting strategy is sufficient to
obtain accurate matrix inverse elements.

%Altogether, we have shown that \pselinv significantly 
%outperforms more general solutions such as the one
%implemented in the \mumps package, and is therefore a
%viable solution even when only diagonal entries of the inverse 
%are needed..

\begin{table}[htbp]
\centering
\begin{tabular}{|c|c|c|}
\cline{2-3}
\multicolumn{1}{c|}{} & shyy161 & stomach \\
\hline
$P$ & $\vert\vert diag(A^{-1}_{\mumps}) - diag(A^{-T}_{\pselinv}) \vert\vert $ & $\vert\vert diag(A^{-1}_{\mumps}) - diag(A^{-T}_{\pselinv}) \vert\vert$ \\
\hline
1 & $4.1813\times 10^{-15}$ & N.A. \\
4 & $4.1837\times 10^{-15}$ & $3.0468\times 10^{-13}$\\
16 & $4.1832\times 10^{-15}$ & $3.0460\times 10^{-13}$ \\
36 & $4.1906\times 10^{-15}$ & $3.0451\times 10^{-13}$ \\
64 & $4.1809\times 10^{-15}$ & $3.0414\times 10^{-13}$\\
\hline
\end{tabular}
\caption{Numerical error of values computed using \pselinv w.r.t. values computed by \mumps 5.0.0 \label{tab.accuracy}}
\end{table}

\section{Conclusion}\label{sec:conclusion}

In this paper, we extend the parallel selected inversion algorithm
called \pselinv, which is originally developed for symmetric matrices, to handle general non-symmetric matrices. The selected
inversion algorithm can efficiently evaluate the elements of $A^{-1}$
indexed by the sparsity pattern of $A^{T}$. From an implementation
perspective, it is more convenient and economical to formulate the selected inversion algorithm to compute selected elements of $A^{-T}$ indexed by the sparsity pattern of $L+U$, where $L,U$ are the $LU$ factors for the possibly permuted matrix of $A$, because such a formulation allows us to overwrite the sparse matrix $L+U$ by the computed elements of $A^{-T}$ \textit{in situ}. We present the data distribution and communication patterns required to perform selected inversion in parallel.  When a large number of processors are used, it is important to exploit coarse-grained level of concurrency available within the elimination trees to achieve high scalability. We also employ a tree-based asynchronous
communication structure for handling various collective communication
operations in the selected inversion algorithm. Our implementation of
\pselinv is publicly available. Our numerical results demonstrates
excellent scalability of \pselinv up to 6400 cores depending on the size
and sparsity of the matrix.  In the near future, we will explore the
efficient implementation of \pselinv on heterogeneous many-core
architecture such as GPU and Intel Knights Landing (KNL).

\section*{Acknowledgment}
This work was partially supported by the Scientific Discovery through
Advanced Computing (SciDAC) program funded by U.S. Department of Energy,
Office of Science, Advanced Scientific Computing Research and Basic
Energy Sciences (M. J., L. L.  and C. Y.), the National Science
Foundation under Grant No. 1450372, and the Center for Applied
Mathematics for Energy Research Applications (CAMERA) (L. L. and C. Y.).  This research used resources of the National Energy Research
Scientific Computing Center, a DOE Office of Science User Facility 
supported by the Office of Science of the U.S. Department of Energy 
under Contract No. DE-AC02-05CH11231.
We thank Volker Blum, Alberto Garc\'{i}a, Xiaoye S. Li, Fran\c{c}ois-Henry
Rouet and Pieter Vancraeyveld for helpful discussion. 

\bibliography{pselinv}

\end{document}